\documentclass{aastex631}

\usepackage{array, boldline, makecell, booktabs}
\usepackage{amsmath}
\renewcommand{\textbf}[1]{#1}

\begin{document}

\title{Polycyclic Aromatic Hydrocarbons (PAHs) as an Extraterrestrial Atmospheric Technosignature}

\correspondingauthor{Dwaipayan Dubey}
\email{ddubey@usm.lmu.de, 2014dwaipayan@gmail.com}

\author[0000-0002-7033-209X]{Dwaipayan Dubey}
\affiliation{Universitäts-Sternwarte, Fakultät für Physik, Ludwig-Maximilians-Universität München, Scheinerstr. 1, D-81679 München, Germany}
\affiliation{Exzellenzcluster `Origins’, Boltzmannstr. 2, D-85748 Garching, Germany}

\author[0000-0002-5893-2471]{Ravi Kopparapu}
\affiliation{NASA Goddard Space Flight Center, 8800 Greenbelt Road, Greenbelt, MD 20771, USA}

\author{Barbara Ercolano}
\affiliation{Universitäts-Sternwarte, Fakultät für Physik, Ludwig-Maximilians-Universität München, Scheinerstr. 1, D-81679 München, Germany}
\affiliation{Exzellenzcluster `Origins’, Boltzmannstr. 2, D-85748 Garching, Germany}

\author[0000-0002-0502-0428]{Karan Molaverdikhani}
\affiliation{Universitäts-Sternwarte, Fakultät für Physik, Ludwig-Maximilians-Universität München, Scheinerstr. 1, D-81679 München, Germany}
\affiliation{Exzellenzcluster `Origins’, Boltzmannstr. 2, D-85748 Garching, Germany}

\begin{abstract}
\textbf{Polycyclic Aromatic Hydrocarbons (PAHs) are prevalent in the universe and interstellar medium but are primarily attributed to anthropogenic sources on Earth, such as fossil fuel combustion and firewood burning. Drawing upon the idea of PAHs as suitable candidates for technosignatures, we investigate the detectability of those PAHs that have available absorption cross-sections in the atmospheres of Earth-like exoplanets (orbiting G-type stars at a distance of 10 parsecs) with an 8m mirror of Habitable Worlds Observatory (HWO). Specifically, we focus on Naphthalene, Anthracene, Phenanthrene, and Pyrene. Our simulations indicate that under current Earth-like conditions, detecting PAH signatures between 0.2–0.515 $\mathrm{\mu m}$ is infeasible. To account for the historical decline in PAH production post-industrial revolution, we explore varying PAH concentrations to assess instrumental capabilities to detect civilizations resembling modern Earth. We also evaluate telescope architectures (6m, 8m, and 10m mirror diameters) to put our results into the context of the future HWO mission. With these four molecules, PAH detection remains infeasible, even at concentrations ten times higher than current levels. While larger mirrors provide some advantages, they fail to resolve the spectral signatures of these molecules with significant signal-to-noise ratios. The UV absorption features of PAHs, caused by $\mathrm{\pi}$-orbital $\rightarrow$ $\mathrm{\pi^*}$-orbital electronic transitions, serve as valuable markers due to their distinct and detectable nature, preserved by the aromatic stability of PAHs. Additional lab measurements are necessary to gather absorption cross-section data beyond UV for more abundant PAHs. This may help further in improving the detectability of these molecules. }

\end{abstract}


\section{Introduction} \label{sec:intro}

The past three decades have been remarkable in exoplanetary science, led by a plethora of exoplanet discoveries and their atmospheric studies. With the current state-of-the-art space and ground-based facilities, nearly 5600 exoplanets have been detected\footnote{\url{https://exoplanetarchive.ipac.caltech.edu/}}, spanning from gas giants to rocky terrestrial planets. Simultaneously, spectroscopy techniques have undergone significant refinement, emerging as a promising avenue for characterizing exoplanet atmospheres, with the key focus on understanding atmospheric pressure-temperature profiles, atmospheric compositions, and its dynamics \citep{2015PASP..127..941C}. Although the initial breakthrough of atmospheric studies has happened on gas giant planets with a lot of follow-up investigations \citep{Charbonneau_2002,2019ApJ...887L..14B,2019NatAs...3.1086T}, the characterization of the atmospheres of rocky planets is slowly coming within our reach \citep{greene2023thermal,lim2023atmospheric,lustig2023jwst,cadieux2024transmission}, offering the exciting opportunity of identifying potentially habitable worlds. Special attention is given to the search for ``biosignatures", i.e. spectral signatures of gases originating from living organisms, offering possible evidence of extraterrestrial life. The concept has earned considerable attention within the exoplanet and astrobiology\footnote{\url{https://www.nationalacademies.org/our-work/astrobiology-science-strategy-for-the-search-for-life-in-the-universe}} communities in recent times. Significant effort is being concentrated on the identification of potential biosignatures as well as the development of strategies to detect them \citep{seager2012astrophysical,grenfell2017review,kaltenegger2017characterize,2018AsBio..18..709C,fujii2018exoplanet,meadows2018exoplanet,schwieterman2018exoplanet,walker2018exoplanet,lammer2019role,o2019expanding}.

Since the first mention by \cite{tarter2006evolution}, the concept of ``technosignatures” has undergone significant development alongside the concept of “biosignatures”. Technosignatures are defined as the manifestations of extraterrestrial technology that could be detected through astronomical observations. The technosignature idea represents a logical extension of the search for biosignatures in the domain of astrobiology, leveraging Earth’s evolutionary life history and technological advancements (NASA Technosignatures Workshop Participants \citeyear{participants2018nasa}). Similar to biosignature investigations, identifying technosignatures involves proposing detectable characteristics inspired by Earth’s perspective and designing strategies to detect them on extraterrestrial worlds. Although the classification and identification of specific technosigantures are still in the infancy stage compared to that of biosignatures \citep{Wright2019Searches,haqq2020observational,lingam2021life}, a significant number of studies on possible technosigantures have paved the way to consider it as one of the important future prospects in the domain of exoplanetary science: waste heat \citep{dyson1960search,2009ASPC..420..415C,wright2014g,kuhn2015global}, artificial illumination \citep{schneider2010far,loeb2012detection,kipping2016cloaking,tabor2021detectability}, artificial atmospheric constituents \citep{owen1980search, Campbell_2005,schneider2010far,2014ApJ...792L...7L,stevens2016observational,kopparapu2021nitrogen,haqq2022detectability}, artificial surface constituents \citep{lingam2017natural}, non-terrestrial artifacts \citep{bracewell1960communications,freitas1980search,rose2004inscribed,haqq2012likelihood}, stellar pollution \citep{shklovsky1966intelligent,whitmire1980nuclear,stevens2016observational}, and megastructures \citep{dyson1960search,arnold2005transit,2013JBIS...66..144F,wright2015g}. Currently, the concept of Search for Extraterrestrial Intelligence (SETI) \citep{wright2018recommendations} has also been developed with a firm belief and expectation to observe Earth’s current technosigantures on planets at interstellar distances with the present technology.

Polycyclic aromatic hydrocarbons (PAHs) are ubiquitous in space, carrying 10–20\% carbon budget of the interstellar medium (ISM) \citep{joblin2011pahs}. \cite{draine2007infrared} and \cite{tielens2008interstellar} provide a more detailed explanation of the contributions of PAHs to the IR emission and their role in the carbon budget of the ISM. They are present in the ISM with a relative number density of 3 $\times$ $\mathrm{10^{-7}}$ respective to hydrogen nuclei \citep{tielens2008interstellar}. From the astrophysical and astrobiological standpoints, PAHs are considered the most interesting complex organic structures for the following reasons: (1) they are crucial in studying the chemical and hydrodynamical evolution of protoplanetary disks and newborn planet atmospheres \citep{gorti2009time,ercolano2022observations,dubey2023polycyclic} as well as in understanding the ionization balance of gaseous atmospheres \citep{thi2019radiation}, and (2) they are anticipated to have significant influence on prebiotic chemistry and abiogenesis, leading an important step toward life formation \citep{ehrenfreund2000organic,ehrenfreund2006experimentally,ehrenfreund2007organics,rapacioli2006formation,wakelam2008polycyclic,sandstrom2011spitzer,2012ApJ...760..120K,puzzarini2017spectroscopic,closs2020prebiotically}. A recent study by \cite{dubey2023polycyclic} shows the formation possibilities of PAHs in the thermalized atmospheres of transiting (irradiated) and directly imaged (non-irradiated) hot-Jupiters. However, their formation processes on terrestrial bodies like Earth and Saturn's moon Titan are still under debate. The detection of PAH molecules on Titan by the space probe Cassini signified the importance of photochemistry for PAH formation (with an abundance of (2 - 3) $\times$ $\mathrm{10^4}$ particles per $\mathrm{cm^3}$) \citep{dinelli2013unidentified,lopez2013large}. PAHs in Earth's atmosphere, on the other hand, stem from various anthropogenic activities ranging from burning wood to incomplete combustion of fossil fuels \citep{ravindra2008atmospheric}. Burning wood could also come from forest fires, which is a natural process. The detection of PAHs could indicate either biological or technological processes, both of which would still imply the presence of life. Disparities in the development status of different countries lead to varying PAH emissions globally, with higher emissions observed in underdeveloped nations \citep{shen2013global}. Developed countries, however, have implemented measures to restrict net PAH emissions since the post-industrial revolution era, reducing the emissions by a factor of almost two.   

Advancing the concept of technosignatures could unveil new avenues by revealing new classes of species further. Theoretical studies have suggested that the detection of $\mathrm{NO_2}$ \citep{kopparapu2021nitrogen} and CFCs (CFC-11 and CFC-12) \citep{haqq2022detectability} on Earth-like exoplanets could serve as potential technosignatures. The presence of PAHs on another planet is an addition to them. Detecting PAHs on exoplanets may provide evidence of industrial activity on their surfaces. In this work, we showcase the study on PAHs as a possible technosignature with the future Habitable Worlds Observatory (HWO) mission concept. We outline our methods in Section \ref{sec:method}. This section explains the global variability in PAH emissions from different sources, molecular cross-sections, and simulation setups for computing the reflected spectra of a planet around a Sun-like star. Our study is confined within the habitable zone of the host star. Section \ref{sec:result} outlines our findings in terms of the detectability of PAH signatures from planet spectra using signal-to-noise ratio estimation. We also discuss about the suitable telescope architecture required to detect PAH with better precision. A concise summary of our result is given in Section \ref{sec:conclusion}.

\section{Methods}
\label{sec:method}

\subsection{Global PAH emission and its variability }
\label{sec:global_variation}

\begin{figure*}
    \centering  
        \begin{minipage}[b]{0.49\columnwidth}
            \includegraphics[width=\columnwidth]{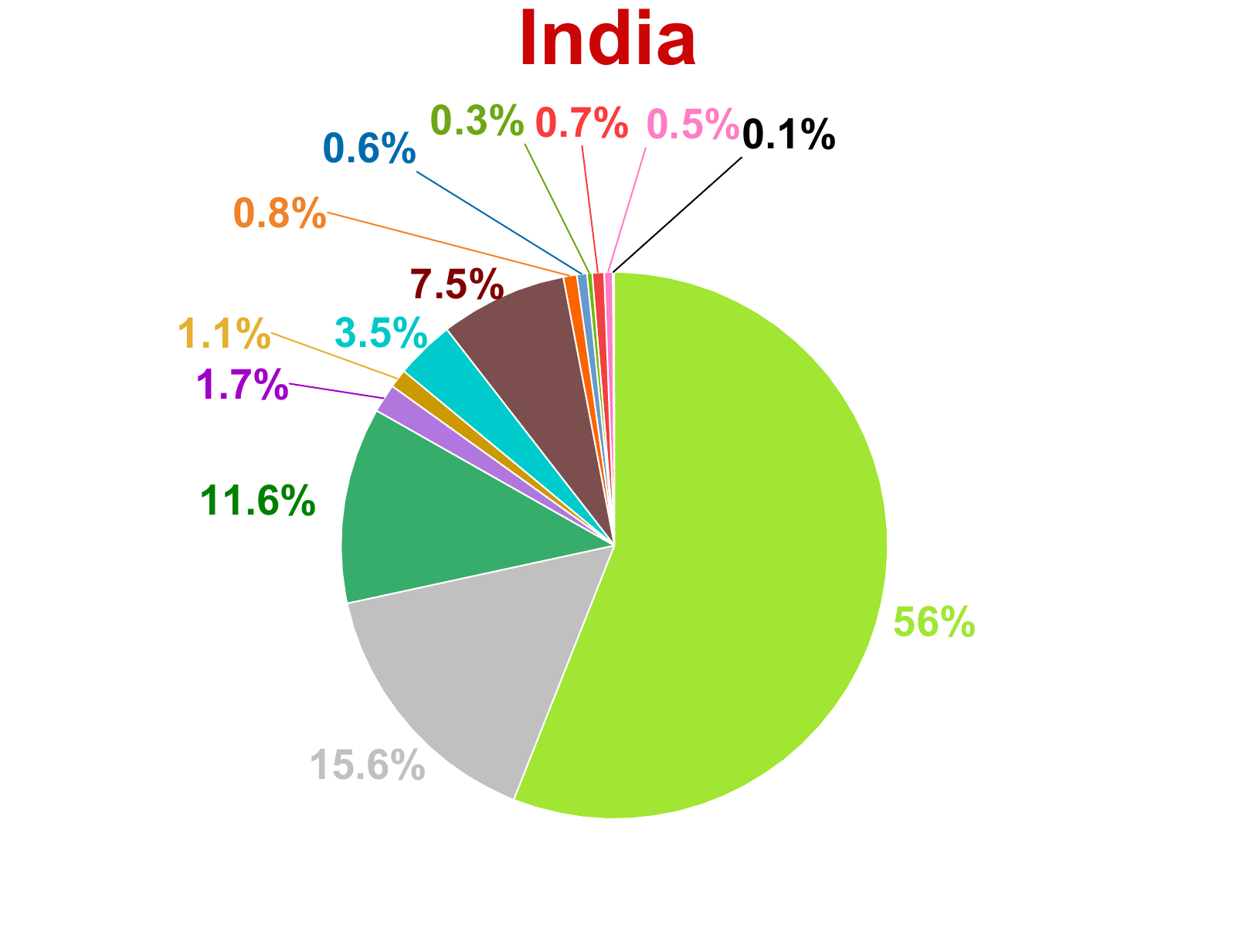}
        \end{minipage}
         \begin{minipage}[b]{0.49\columnwidth}
            \includegraphics[width=\columnwidth]{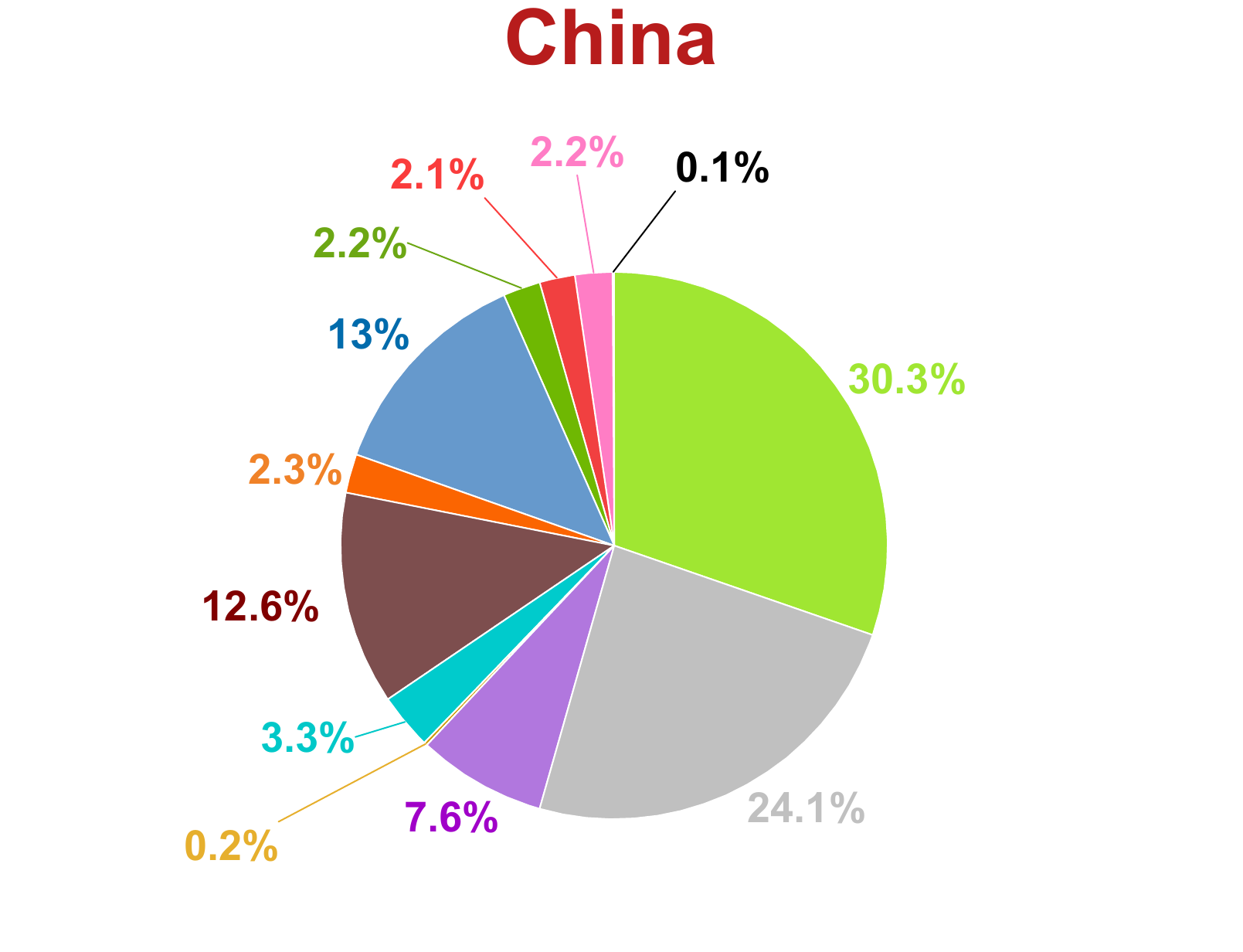}
        \end{minipage}
         \begin{minipage}[b]{0.49\columnwidth}
            \includegraphics[width=\columnwidth]{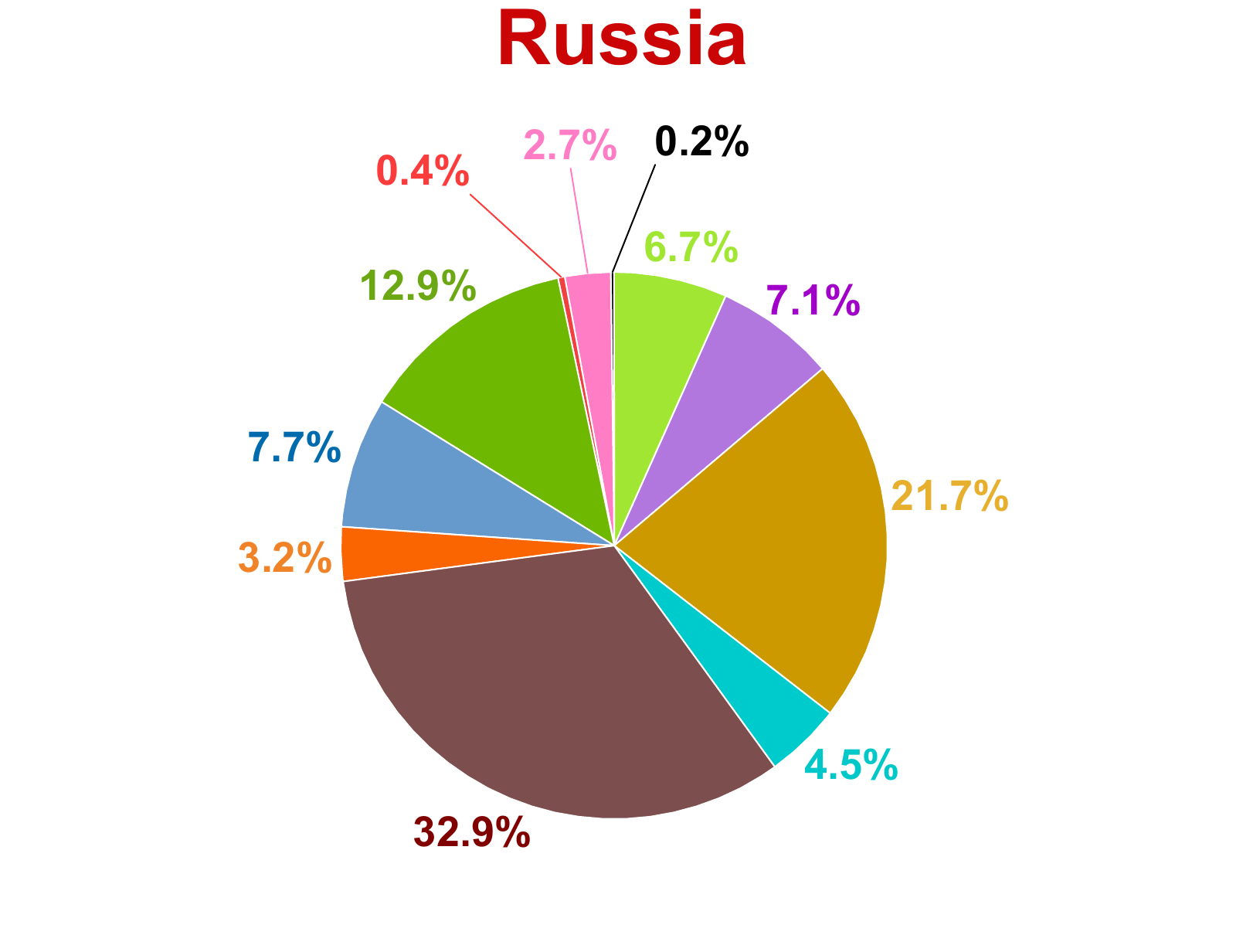}
        \end{minipage}
         \begin{minipage}[b]{0.49\columnwidth}
            \includegraphics[width=\columnwidth]{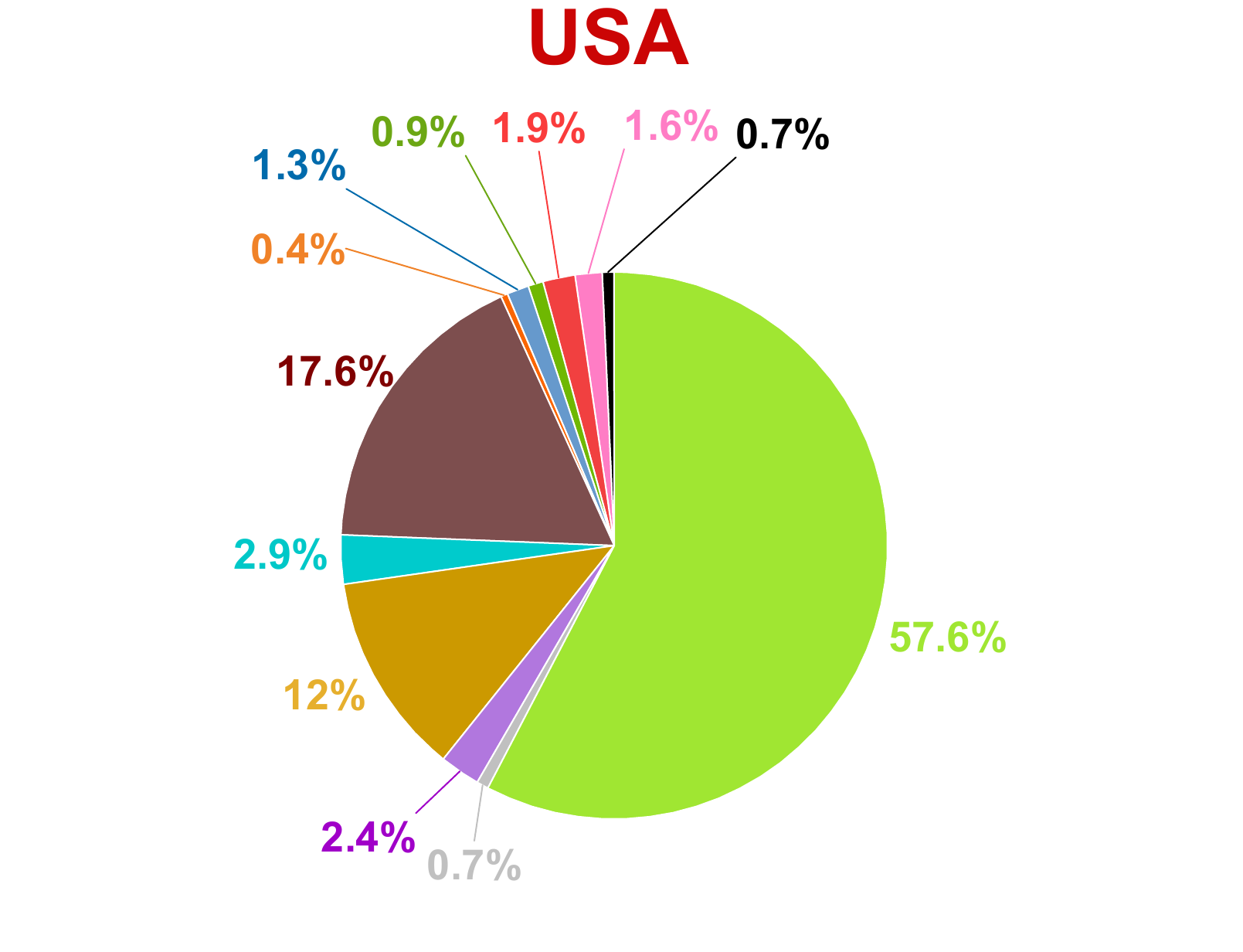}
        \end{minipage}
         \begin{minipage}[b]{0.49\columnwidth}
            \includegraphics[width=\columnwidth]{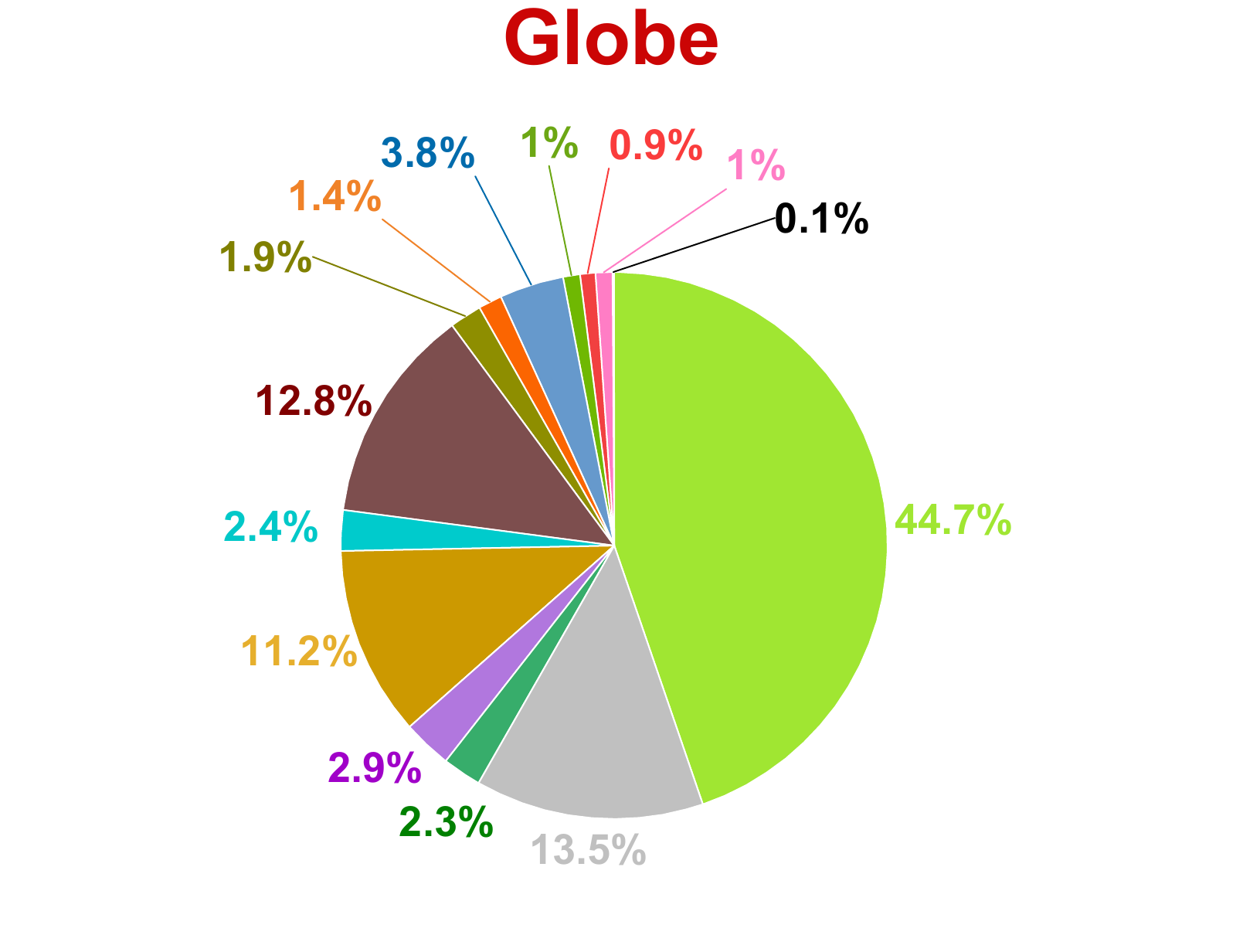}
        \end{minipage}
        \begin{minipage}[b]{0.49\columnwidth}
            \includegraphics[width=\columnwidth]{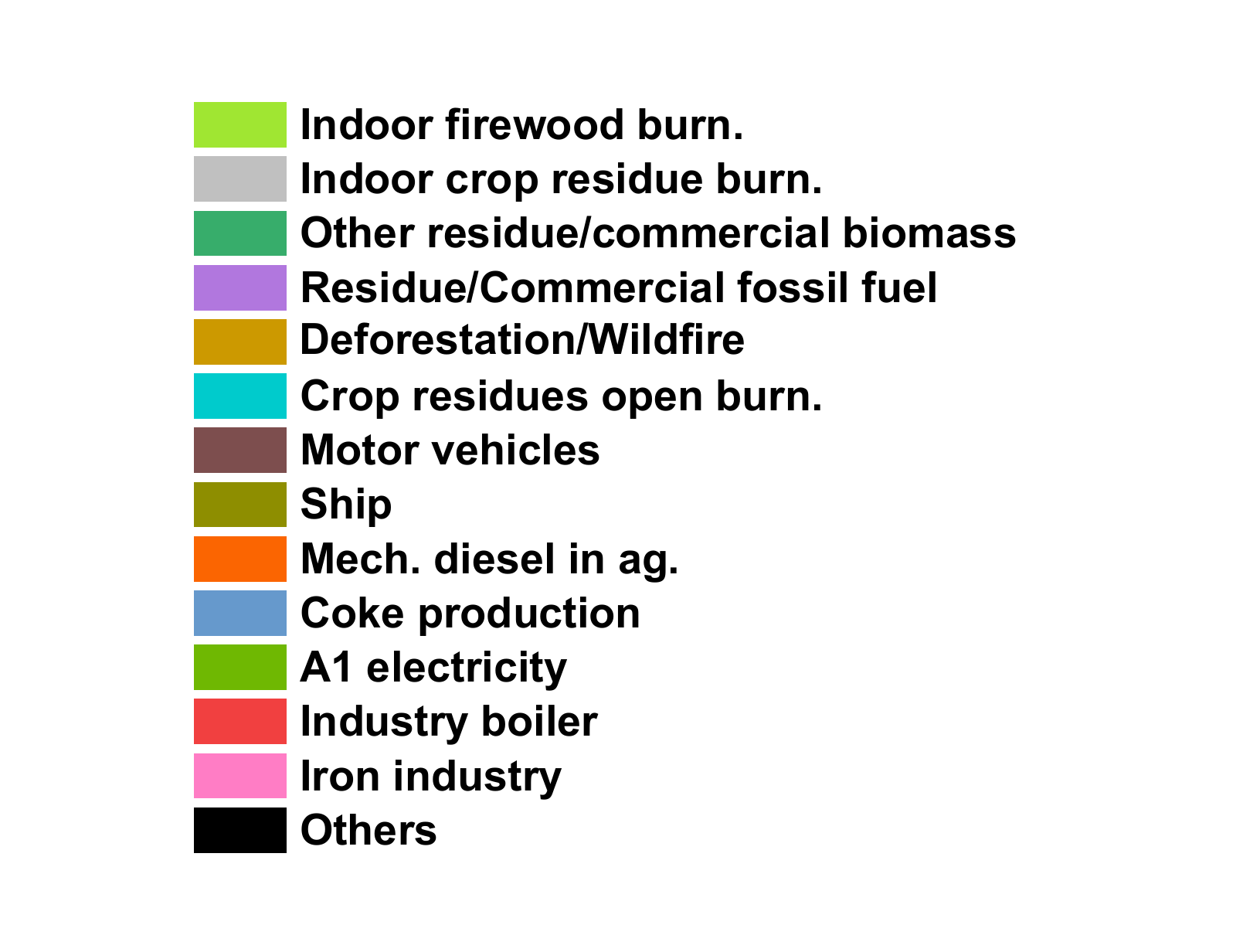}
        \end{minipage}
\caption{Contributions of different anthropogenic sources to the net global PAH emission including four major PAH producing countries in the year 2007 \citep{shen2013global}. The pie charts illustrate the total emissions for each country as a whole. These countries exemplify the conditions found in both developed and developing regions globally. Firewood and crop residue burn, deforestation/wildfire, and motor vehicles are the dominating source profiles across the globe.} 
\label{fig:pie}
\end{figure*}

PAHs are very crucial in the context of Earth due to their significant contribution to global pollution and human health hazards \citep{laflamme1978global,perera1997environment}. Based on the PKU-FUEL-2007 database \citep{wang2013high}, \cite{shen2013global} calculated the historical and future trends in country-specific PAH emissions from 1960 to 2030. We obtained their data on global PAH emissions, including specific data from four major contributing countries: India, China, Russia, and the US (see Figure \ref{fig:pie}). While India and China represent the developing countries, Russia and the US provide an understanding of the developed regions. Figure \ref{fig:pie} illustrates the relative contributions of different anthropogenic sources to the net PAH production for them. The variability in PAH source profiles across different countries is particularly attributed to their different developmental position, energy infrastructure, and vegetation coverage. Globally, including India, China, and the US, indoor firewood burning is the major contributor to the net PAH emission. While it is one of the major energy sources in rural areas of India (90\%) and China (70\%) \citep{balakrishnan2011air,zhou2017comprehensive}, the situation in the US differs significantly, with the prevalence of certified wood stoves in rural households (U.S. Environmental Protection Agency \citeyear{osti_5473300}). Russia, on the contrary, is mostly dominated by the emissions from motor vehicles. The coke production sector (extraction of carbon-rich fuel from coal which is further used in the steel-making industry and other industries) has historically held significance in Russia and China, yet it is presently declining due to the discontinuation of the use of beehive coke ovens. Deforestation is also quite significant for Russia and the US.

For this study, we have selected four PAHs: Naphthalene ($\mathrm{C_{10}H_8}$), Anthracene ($\mathrm{C_{14}H_{10}}$), Phenanthrene ($\mathrm{C_{14}H_{10}}$), and Pyrene ($\mathrm{C_{16}H_{10}}$). Our selection criteria are twofold: firstly, these compounds represent the simplest PAHs ranging from those composed of 2 benzene rings to 4 benzene rings (easily formed and more abundant); secondly, their UV cross sections are publicly available in the community. Measured global emissions (in Mg year$^\mathrm{{-1}}$ unit) of these PAHs in 2007 are shown in Table \ref{tab:density}. For further calculations, we utilize the following mathematical prescription to convert the PAH emissions to volume mixing ratio unit (parts per billion (ppb)). We have made two key assumptions relevant to this analysis: (1) Emissions are uniformly distributed within a defined volume of air in the atmosphere. Since emissions occur close to the ground, we assume that the PAH mass is concentrated within the first 5 to 10 km of altitude, primarily within the troposphere, where these heavier molecules are likely more prevalent. This region encompasses most of the atmospheric mass. (2) \textbf{According to \cite{shen2013global}, the global trend in PAH emissions has remained relatively consistent over the past decades since 1960. However, the atmospheric lifetime of Naphthalene, Anthracene, Phenanthrene, and Pyrene on Earth is on the order of hours to a few days (Naphthalene: \cite{el2005toxicological}; Anthracene: \cite{shahsavar2023atmospheric}; Phenanthrene: \cite{zhao2016atmospheric}, and Pyrene: \cite{national1983atmospheric} (National Research Council)). A significant portion of these molecules is expected to be degraded through reactions with reactive species such as OH radicals, ozone, and other oxidizing agents. Recombination processes are relatively slower than their destruction mechanisms, leading to a net decrease in atmospheric concentrations over time. To account for this loss mechanism, we have incorporated a decay term into the PAH concentration calculations (see Equation \ref{eq2}). Given the absence of a detailed PAH kinetic network tailored for an Earth-like atmosphere, this approach represents the most realistic estimation of present-day PAH concentrations.}

The Earth's radius is approximately 6371 km. By considering an atmospheric shell extending 10 km above the Earth's surface, where emissions are assumed to be present, the volume of this atmospheric shell can be calculated as follows:

\begin{equation}\label{eq1}
    \mathrm{V_{air}} = \mathrm{\frac{4}{3}\pi (R^3_{outer} - R^3_{inner})} \approx 5.11 \times \mathrm{10^{18}} \text{ $\mathrm{m^3}$}
\end{equation}

\noindent where $\mathrm{R_{inner}}$ =  Earth radius $\approx$ 6371 km and $\mathrm{R_{outer}}$ = Earth radius with the atmospheric shell $\approx$ 6381 km. 

\textbf{If the emission rate of an individual PAH is measured to be F Mg each year (specific values provided in Table \ref{tab:density}), then, based on assumption 2 (i.e., the global emission rate has remained consistent since 1960), the remaining atmospheric density of PAHs ($\mathrm{C_{g/m^{3}}}$ [now]) at the current times can be calculated using Equation \ref{eq2}:}


\begin{equation}\label{eq2}
    \mathrm{C_{g/m^{3}}} [\text{now}]= \sum_{i=1960}^{2024}\Big[\mathrm{\frac{F_i}{5.11 \times 10^{18}} \times 10^{6}\Big]e^{-(2024-i)/\tau}} \approx \mathrm{\frac{F_{2024}}{5.11 \times 10^{18}} \times 10^{6}} \approx \mathrm{\frac{F_{2007}}{5.11 \times 10^{18}} \times 10^{6}}
\end{equation}

\textbf{In Equation \ref{eq2}, the concept of atmospheric lifetime is applied, analogous to an exponential decay process. Here, $\tau$ represents the lifetime of the specific molecules, typically ranging from hours to a few days, during which they are degraded through chemical reactions and other loss mechanisms. Suppose the atmospheric lifetime of a molecule is x days (= $\mathrm{\frac{x}{365}}$ years). In that case, the exponential term in the decay process will lead to the near-complete depletion of PAH concentrations from previous years, as the exponent becomes largely negative over time. Therefore, focusing on the PAH concentration for the current year provides the most realistic estimate for assessing the detectability.}

At standard temperature and pressure (STP: 0$^{\circ}$C and 1 atm), 1 mole of any ideal gas occupies 22.4 liters of volume = 0.0224 $\mathrm{m^3}$. Therefore, 1 $\mathrm{m^3}$ of air contains $\mathrm{\frac{1}{0.0224}}$ $\approx$ 44.64 moles of air. Now, let's say the molecular mass of the PAH is M g/mol. We can convert the density from \textbf{$\mathrm{C_{g/m^{3}}}$ [now]} to $\mathrm{C_{moles/m^{3}}}$ using Equation \ref{eq3}:

\begin{equation}\label{eq3}
    \mathrm{C_{moles/m^{3}}} = \mathrm{\frac{C_{g/m^{3}} [\text{now}]}{M}}
\end{equation}

\noindent Finally, the volume mixing ratios in ppb are obtained by:

\begin{equation}\label{eq4}
    \text{volume mixing ratio (ppb)} = \mathrm{\frac{C_{moles/m^{3}}}{44.64}} \times 10^{9}
\end{equation}

\noindent \textbf{These steps led to a global average volume mixing ratio of 0.0078 ppb for Naphthalene, 0.00024 ppb for Anthracene, 0.0014 ppb for Phenanthrene, and 0.00041 ppb for Pyrene.}
\begin{table*}
\centering
\caption{Worldwide net emission (Mg year$^\mathrm{{-1}}$) of selected PAHs in 2007.}
\begin{tabular*}{\columnwidth}{@{\extracolsep{\fill}}lccc@{}}

\hline
\hline
\noalign{\smallskip}
PAHs & Global Net Emission & PAHs & Global Net Emission\\
 & (Mg year$^\mathrm{{-1}}$) &  & (Mg year$^\mathrm{{-1}}$)\\
\noalign{\smallskip}
\hlineB{3.5}
\noalign{\smallskip}
Naphthalene (Nap) & 2.3 $\times$ $\mathrm{10^5}$ & Phenanthrene (Phe) & 5.3 $\times$ $\mathrm{10^4}$ \\
Anthracene (Ant) & 1.0 $\times$ $\mathrm{10^4}$ & Pyrene (Pyr) & 1.9 $\times$ $\mathrm{10^4}$ \\
\noalign{\smallskip}
\hline
References & \multicolumn{3}{c}{\cite{shen2013global}} \\
\noalign{\smallskip}
\hline
\hline
\label{tab:density}
\end{tabular*}
\end{table*}

\subsection{Molecular cross-sections}
\label{sec:cross_section}

\begin{figure*}
    \centering  
        \begin{minipage}[b]{0.49\columnwidth}
            \includegraphics[width=\columnwidth]{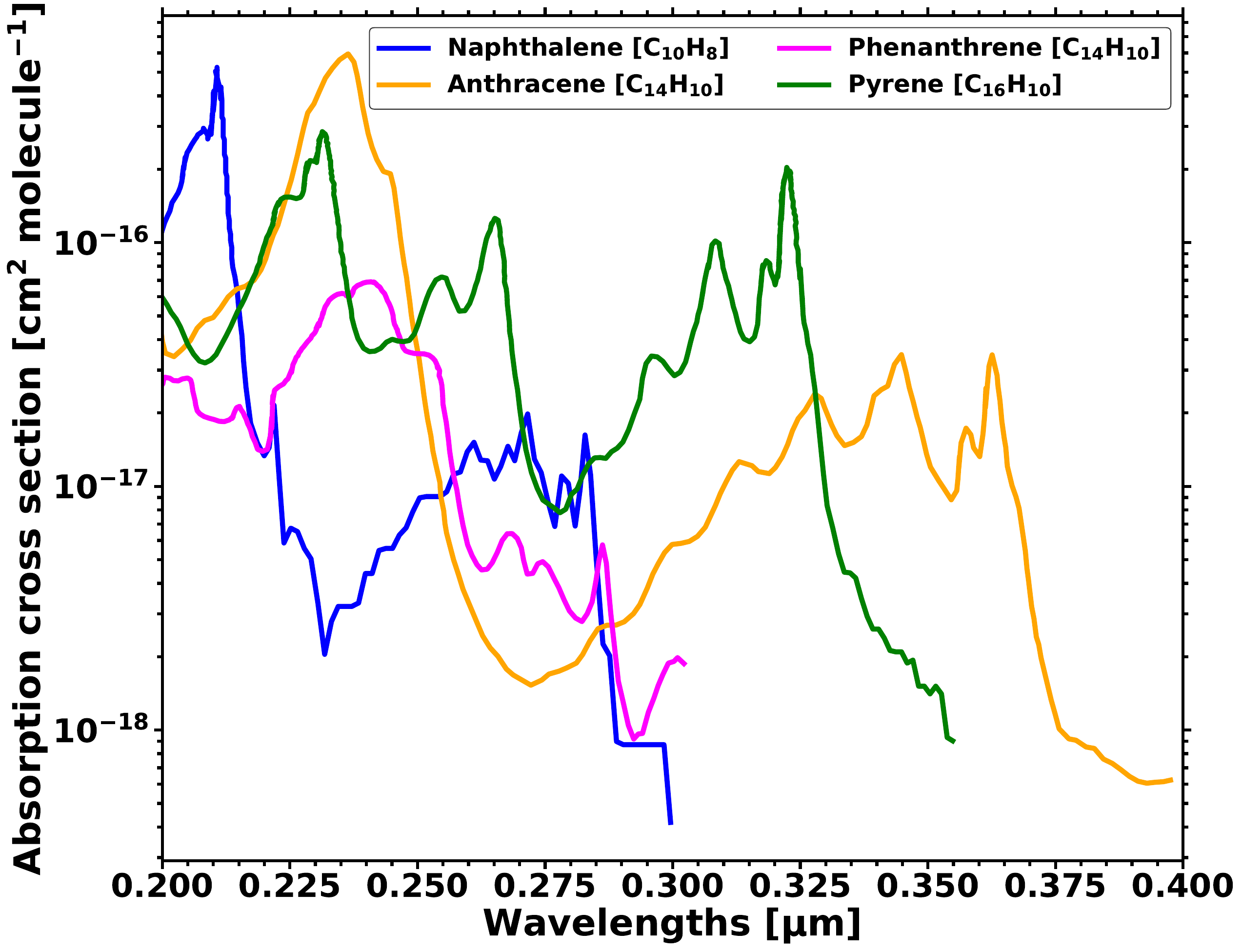}
            \begin{center}
      \textbf{(a)}
    \end{center}
        \end{minipage}
         \begin{minipage}[b]{0.49\columnwidth}
            \includegraphics[width=\columnwidth]{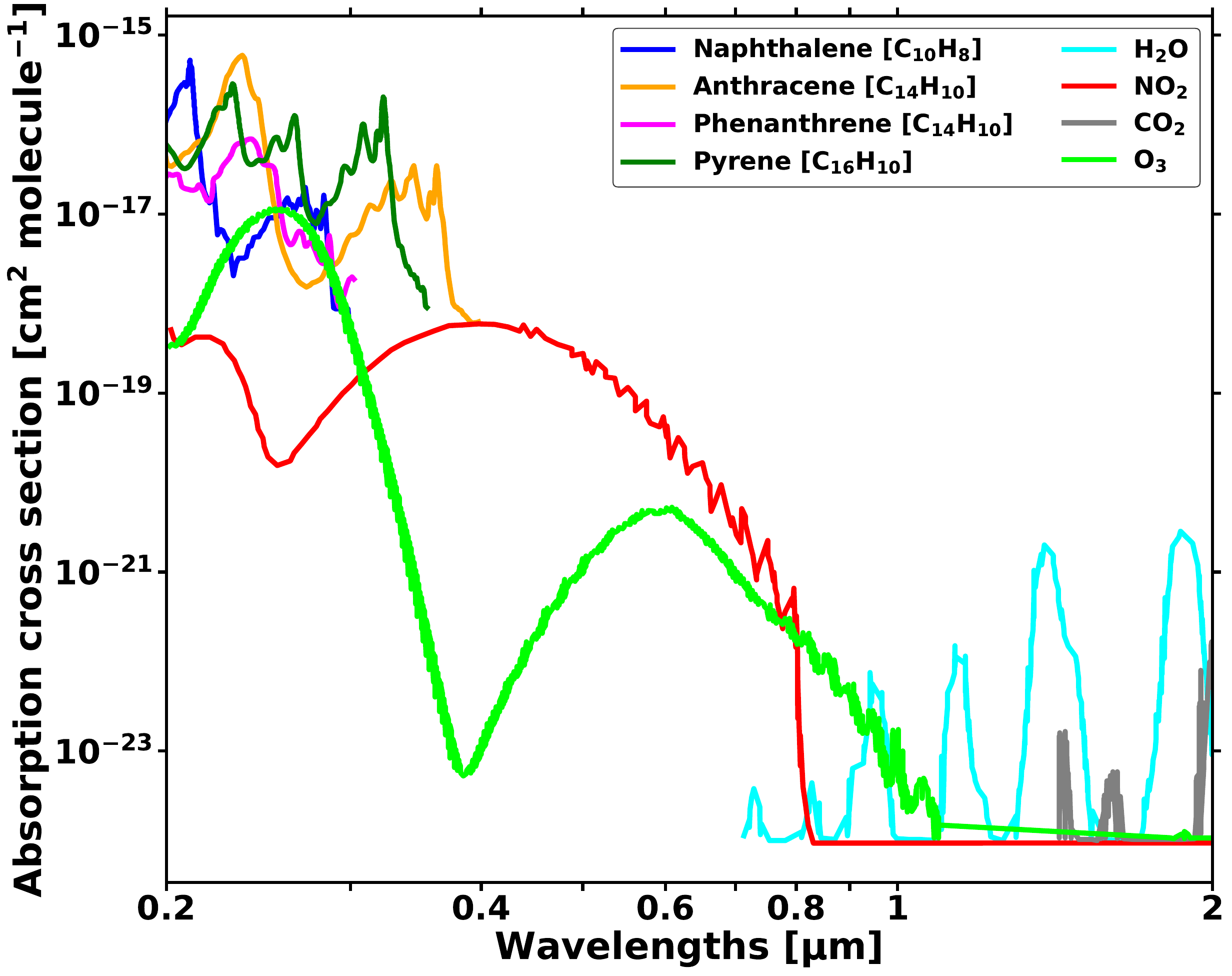}
            \begin{center}
      \textbf{(b)}
    \end{center}
        \end{minipage}
\caption{(a) Cross sections of four different PAHs in the UV region (0.2-0.4 $\mathrm{\mu m}$). On average, Naphthalene, Anthracene, and Pyrene have stronger cross-sections among all 4 considered. (b) Collective cross sections of PAHs and other molecules on Earth (taken from \cite{kopparapu2021nitrogen}) for the wavelength coverage of a nominal 8m class Habitable Worlds Observatory (0.2-2 $\mathrm{\mu m}$). 
PAHs have stronger features than other molecules in this region.
The region between 0.3-0.4 $\mathrm{\mu m}$ serves as a promising window to detect PAHs due to the least overlap with other molecular cross-sections.}
\label{fig:cross_section}
\end{figure*}

Molecular cross-sections are essential for radiative transfer calculations. For Earth-like planets, we have considered several key molecules: $\mathrm{H_2O}$, $\mathrm{NO_2}$, $\mathrm{CO_2}$, $\mathrm{O_3}$, $\mathrm{O_2}$ along with the four PAHs mentioned in Section \ref{sec:global_variation}. Figure \ref{fig:cross_section} illustrates the absorption cross-sections of these molecules. Due to its lower cross-section value, $\mathrm{O_2}$ is not depicted here. The PAH cross-section data are taken from the MPI-Mainz UV/VIS Spectral Database\footnote{\url{https://uv-vis-spectral-atlas-mainz.org/uvvis/cross_sections/}} (Naphthalene: \cite{grosch2015uv, suto1992quantitative}, Anthracene: \cite{thony1997gas}, Phenanthrene: \cite{kitagawa1968absorption}, Pyrene: \cite{thony1997gas}). Cross-sections of other key gaseous species are obtained from \cite{kopparapu2021nitrogen}. It is noteworthy that the strength of the PAH cross-sections is significantly higher than other molecules.  While there is a slight overlap between $\mathrm{O_3}$ and PAH cross-sections in the near-UV region, the range from 0.3-0.4 $\mathrm{\mu m}$ presents a promising window for PAH detection as a technosignature.

\subsection{Radiative transfer calculation and simulation setup}
\label{sec:radiative_transfer}

To estimate the detection significance of PAHs on Earth-like exoplanets, we incorporate the molecular volume mixing ratios (VMRs) into the radiative transfer tool, Planetary Spectrum Generator (PSG\footnote{\url{https://psg.gsfc.nasa.gov/index.php}}: \cite{villanueva2018planetary,villanueva2022fundamentals}), to simulate reflected light spectra from the planet surface and atmosphere. PSG utilizes the latest spectroscopic parameterizations and does a spherical correction to plane-parallel geometry, using the radiative transfer computations of a plane-parallel wave. Aerosols were not included in our model, and we used the default 2-stream approximation for radiative transfer. It has the potential to simulate synthetic observations for any given telescope architecture and instrumental parameters (wavelength coverage, resolution, different noise parameters, etc.). Due to the lack of data on the vertical distribution of PAH on Earth, we have maintained their VMRs constant across all pressure layers. On the other hand, for $\mathrm{H_2O}$, $\mathrm{NO_2}$, $\mathrm{CO_2}$, $\mathrm{O_3}$, $\mathrm{O_2}$, modern Earth vertical profiles have been considered during spectral calculation.

\begin{figure*}[b]
    \centering  
        \begin{minipage}[b]{0.49\columnwidth}
            \includegraphics[width=\columnwidth]{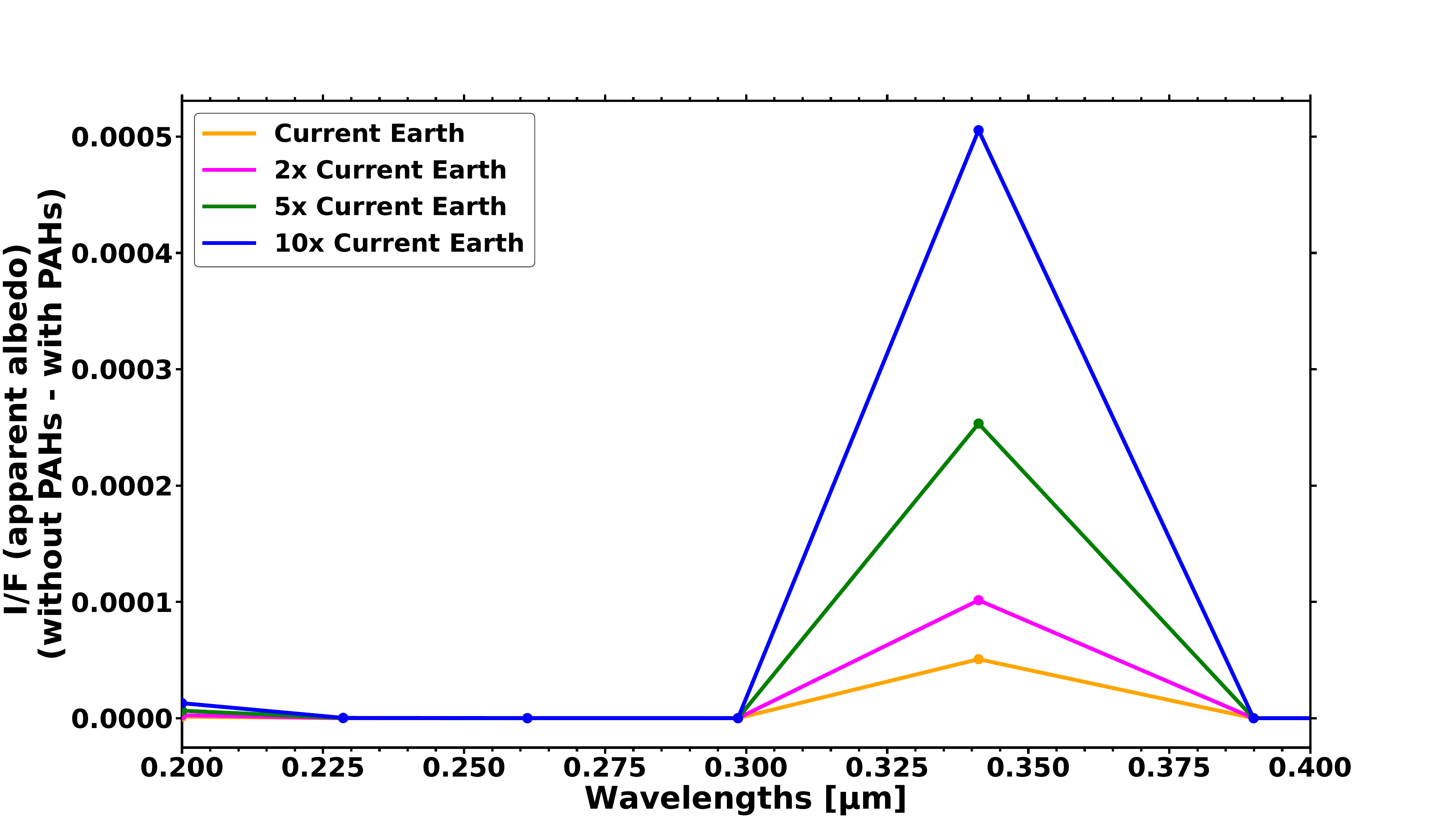}
         \begin{center}
             \textbf{(a)}
        \end{center}
        \end{minipage}
         \begin{minipage}[b]{0.49\columnwidth}
            \includegraphics[width=\columnwidth]{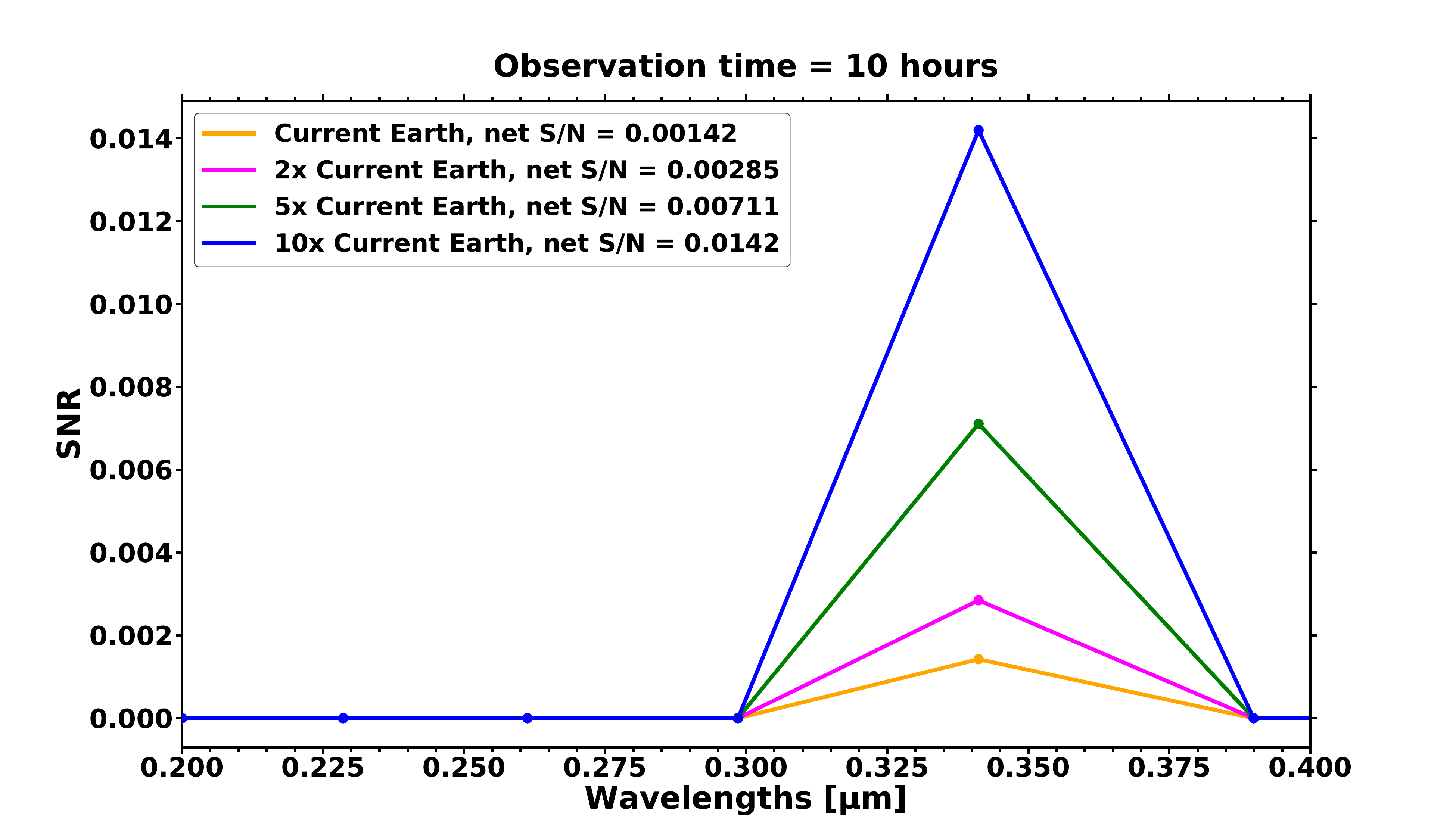}
             \begin{center}
                \textbf{(b)}
             \end{center}
        \end{minipage}
         \begin{minipage}[b]{0.49\columnwidth}
            \includegraphics[width=\columnwidth]{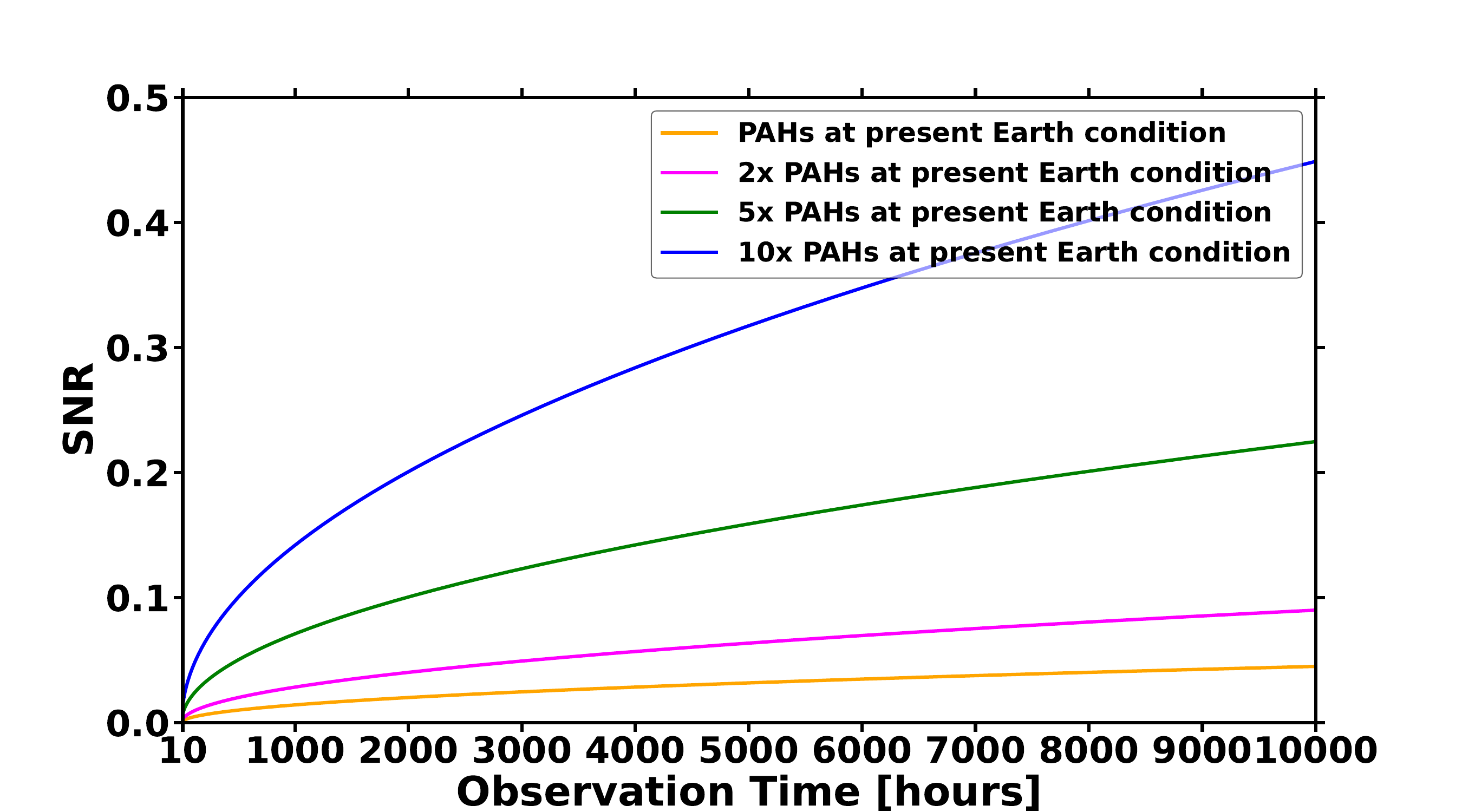}
            \begin{center}
                \textbf{(c)}
             \end{center}
        \end{minipage}
\caption{(a) Contrast difference with and without PAHs in the reflected spectra for an Earth-like exoplanet around a Sun-like star for different levels of PAHs in the planet atmosphere (\textit{orange}: current Earth, \textit{magenta}: 2x current Earth, \textit{green}: 5x current Earth, \textit{blue}: 10x current Earth). This is calculated for the 8m mirror of the HWO telescope. The circles represent the wavelength bin points of the spectra. The difference in the 0.2-0.225 $\mathrm{\mu m}$ is due to Naphthalene and the difference in the 0.3-0.38 $\mathrm{\mu m}$ is due to the cumulative effect of Anthracene and Pyrene. Different legends represent the different concentrations of PAHs, scaled by a factor, on Earth's atmosphere. (b) Calculated signal-to-noise ratio (SNR) for different PAH concentrations as a function of wavelength. It is done for a future HWO-like telescope for an observation time of 10 hours and a planet 10 pc away from Earth. In legend, the net SNR is calculated following Equation \ref{eq5}. Similar to Figure \ref{fig:spectra_different_earth_condition}a, the circles represent the wavelength bin points of the SNR. (c) Measured net SNR value as a function of total observation time using an HWO-like telescope. \textbf{With 10 times the current Earth concentration of Naphthalene, Anthracene, Phenanthrene, and Pyrene, achieving a significant SNR is unachievable with the current setup.}}
\label{fig:spectra_different_earth_condition}
\end{figure*}

Our study focuses on Earth-like planets revolving around G-type stars at a distance of 10 parsecs from Earth (this specific configuration is to study Earth as an exoplanet and 10 pc is 
a standard distance used for such kinds of studies). We consider observations when the planet is at quadrature (90$^{\circ}$ or 270$^{\circ}$) to maximize the angular separation between the star and the planet, thereby improving the ability to resolve the planet. Our study is conducted with a focus on a future observatory similar to the HWO mission concept, utilizing the ECLIPS (Extreme Coronograph for Living Planetary Systems)\footnote{\url{https://www.luvoirtelescope.org/}, Section 1.11.2, Page 75} coronagraph. This instrument will be capable of resolving exoplanet observation for a wavelength window of 0.2-2 $\mathrm{\mu m}$ with its three different channels: near-ultraviolet band (NUV, 0.2-0.525 $\mathrm{\mu m}$), visible band (VIS, 0.515-1.03$\mathrm{\mu m}$), and near-infrared band (NIR, 1-2$\mathrm{\mu m}$). The NUV channel is primarily designed for high-contrast imaging, offering a maximum resolution of R $\sim$ 7. Meanwhile, the VIS and NIR channels are equipped with both an imaging camera and an integral field spectrograph, operating at resolutions of R $\sim$ 140 and R $\sim$ 70, respectively. We investigate the detectability of PAHs for different levels of current Earth-like concentrations using the signal-to-noise ratio (SNR) calculation. Additionally, we follow up our analysis with different instrumental architectures  (mirror diameters: 6m, 8m, and 10m) to further comprehend the detectability of PAHs. To estimate differential SNR, we follow the scheme from \cite{Lustig-Yaeger_2019}:

\begin{equation}\label{eq5}
    \mathrm{SNR = \sqrt{\sum_{i=0}^{n}\Big(\frac{dS_i}{N_i}\Big)^2} \quad\text{, where dS = $\mathrm{S_1 - S_0}$}}
\end{equation}

\noindent Here, $\mathrm{S_1}$ = planet spectra with all other molecules except PAHs, $\mathrm{S_0}$ = planet spectra with all molecules including PAHs, N = simulated noise for observation, and n = total count of wavelength bins.  

Total noise for a specific observation is calculated as a collective sum of noises from the detector, source, background, and telescope (read noise and dark current) (discussed in \cite{villanueva2018planetary,villanueva2022fundamentals} and PSG website). PSG demonstrates strong compatibility with other noise models when calculating photon count rates and spectral precision, showcasing a good fit within the broader framework (see the Appendix of \cite{kopparapu2021nitrogen}). 


\section{Detectability of PAHs}
\label{sec:result}

\subsection{Different level of atmospheric PAH concentration}
\label{sec:different_level_pah}

From Table \ref{tab:density}, we have estimated the global volume mixing ratio values for each PAH. Figure \ref{fig:spectra_different_earth_condition}a illustrates the difference in planetary spectra between atmospheres with and without PAHs, as observed with the HWO telescope. It also demonstrates how the spectra would vary if the planet inherited different levels of PAHs in its atmosphere. \textbf{Considering the historical decline in global PAH production post the industrial revolution era, we have adopted twice the concentrations from Section \ref{sec:global_variation} as the current Earth PAH level for our analysis. This assessment is based on the uncertainty outlined in \cite{shen2013global} (See Figure 4a), which represents the best-case scenario for this context.} We have scaled this PAH abundance by factors of 2x, 5x, and 10x relative to the current Earth level. An enhancement in PAH abundance on the planet would significantly influence the resultant spectra, increasing the strength of the absorption signal by several orders of magnitude. Within the 0.2-0.225 $\mathrm{\mu m}$ range, Naphthalene primarily accounts for the PAH signal owing to its higher order cross-section compared to others. Conversely, Anthracene and Pyrene exert greater influence on the PAH signal between 0.3 and 0.4 $\mathrm{\mu m}$. One significant limitation of our study is that we are restricted to UV observations (0.2-0.4 $\mathrm{\mu m}$) with the instrument. As discussed in Section \ref{sec:global_variation}, this caveat arises from the lack of available PAH cross-sections across the entire wavelength range of the instrument. Having complete cross-sections would allow for a more accurate inference of the overall differences between the planet's spectra.

In Figure \ref{fig:spectra_different_earth_condition}b, we present the SNR as a function of wavelength for different Earth-like scenarios as described previously. This analysis is conducted for a future 8m HWO observation with an exposure time of 10 hours. The resulting net SNR shown in this figure is calculated following Equation \ref{eq5}. It is important to notice here that the overall SNR increases almost linearly with the increase in the PAH level in the atmosphere. However, a proper argument would need the support of a proper retrieval assessment. Even with 10 hours of observation and a PAH concentration 10 times higher than that of the current Earth, the features are not detectable with a plausible SNR value (SNR value is completely within the noise of the simulated observation). Therefore, additional observation time is necessary to resolve PAH fingerprints in the spectra. Figure \ref{fig:spectra_different_earth_condition}c showcases the trend of detected SNR with variable observation time using the future HWO-like telescope. \textbf{Even with higher PAH concentrations, 2x, 5x, and 10x PAH amounts fail to achieve a significant SNR even after 10000 hours of observation.} Furthermore, it is practically unfeasible to observe atmospheres for such extended periods of time. This emphasizes the need for PAH cross-sections in the visible and NIR regions. With complete cross-sections, the accuracy of observations could be greatly enhanced, reducing the required observation time to detect a significant PAH signal.

\begin{figure*}[h!]
    \centering  
        \begin{minipage}[b]{0.49\columnwidth}
            \includegraphics[width=\columnwidth]{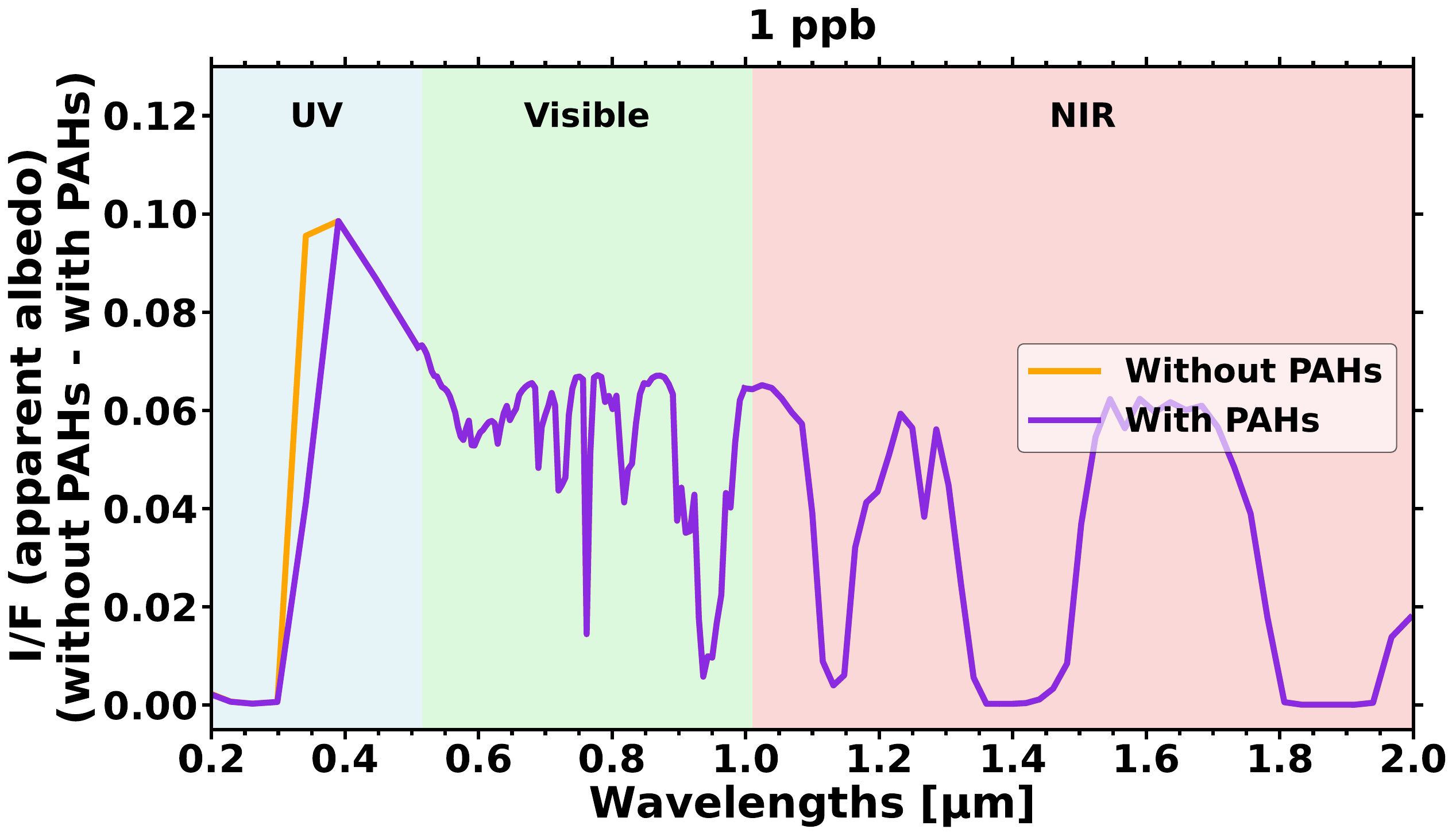}
        \end{minipage}
         \begin{minipage}[b]{0.49\columnwidth}
            \includegraphics[width=\columnwidth]{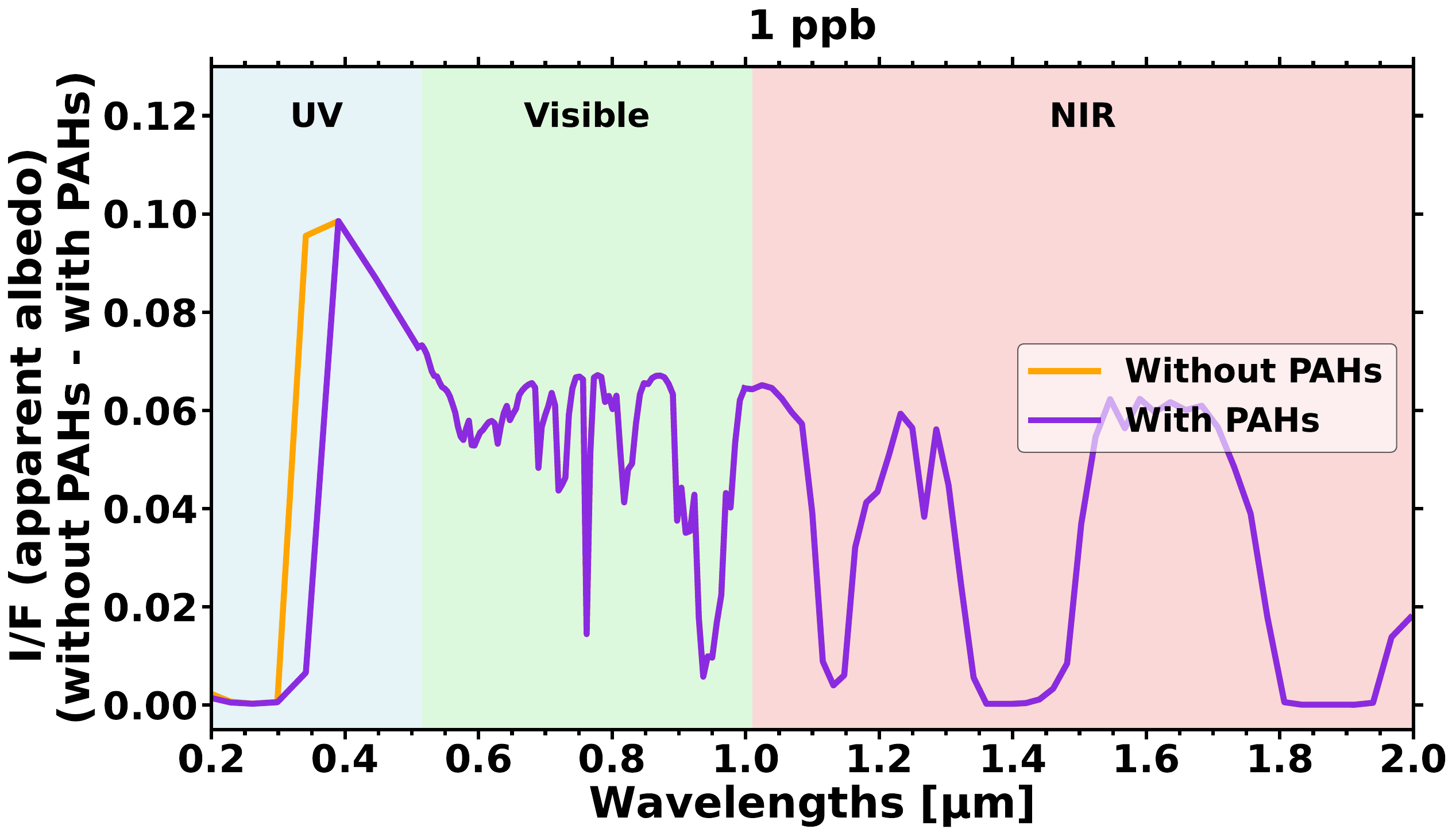}
        \end{minipage}
\caption{Complete spectrum of an Earth-like exoplanet with different PAH concentrations around a Sun-like star using HWO. \textit{Left panel}: 1 ppb concentration for all PAHs. \textit{Right panel}: 10 ppb concentration for all PAHs. The wavelength coverage of HWO is divided into three different channels with different resolutions - \textit{Blue band}: UV (R $\sim$ 7), \textit{Green band}: Visible (R $\sim$ 140), and \textit{Red band}: NIR (R $\sim$ 70). The differences between the spectra are only limited to the UV region (0.2-0.4 $\mathrm{\mu m}$) due to the unavailability of PAH cross-sections.}
\label{fig:spectra_different_ppb}
\end{figure*}

\textbf{A significant limitation of this study is that the analysis is restricted to four PAH species (Naphthalene, Anthracene, Phenanthrene, and Pyrene) due to the availability of absorption cross-sections. These molecules, though assumed to have lifetimes of a few days in the atmosphere, are not completely degraded under typical conditions, as their destruction requires strong UV radiation. Instead, their lifetimes are primarily influenced by reactions with reactive radicals, leading to the formation of substituted PAHs rather than their complete removal. Consequently, PAHs persist in the atmosphere in various stable forms. To accurately capture PAH signatures and their detectability, it is essential to study a broader range of molecules within the PAH family, as their collective behavior provides a more comprehensive understanding.} Given the wide variety of PAHs present in Earth's atmosphere, this advancement could position PAHs as a key technosignature to consider in the near future (see Appendix \ref{sec:future} for specific PAHs to be studied for cross-sections). A complementary analysis is presented in Figure \ref{fig:spectra_different_ppb} for planets that acquire higher concentrations (1 ppb and 10 ppb) for all PAHs. This figure illustrates how the difference in planetary spectra would appear when the PAH concentrations are set at 1 and 10 ppb levels for all PAHs in the atmosphere. A distinct PAH signal is discernible for both scenarios, characterized by a pronounced rise in absorption features as the PAH concentration increases from 1 ppb to 10 ppb.

\subsection{Telescope architecture}
\label{sec:architecture}

Furthermore, we have conducted an in-depth analysis of critical telescope size for the upcoming HWO mission to gain deeper insights into our previous findings and to propose the most suitable telescope architecture for PAH detection. The diameter of the telescope determines the incoming flux and consequently, the signal received from the planet. This parameter needs to be optimized to strike a balance between incoming background noise and signals from the planet's surface. With the current HWO concept of an 8m telescope diameter, we have explored two alternative diameter options of 6m and 10m. 

\begin{figure*}[h!]
    \centering  
        \begin{minipage}[b]{0.49\columnwidth}
            \includegraphics[width=\columnwidth]{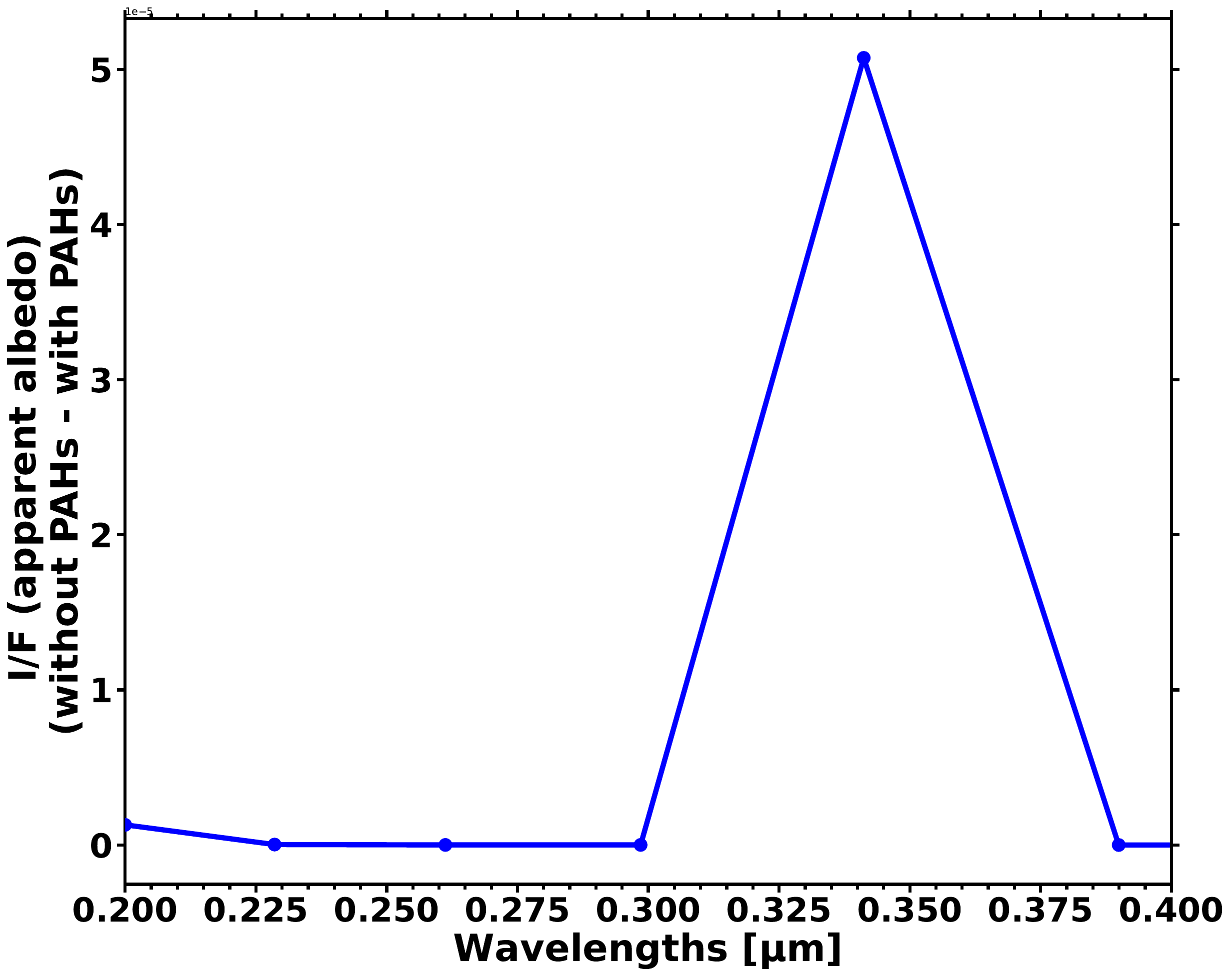}
        \begin{center}
             \textbf{(a)}
        \end{center}
        \end{minipage}
         \begin{minipage}[b]{0.49\columnwidth}
            \includegraphics[width=\columnwidth]{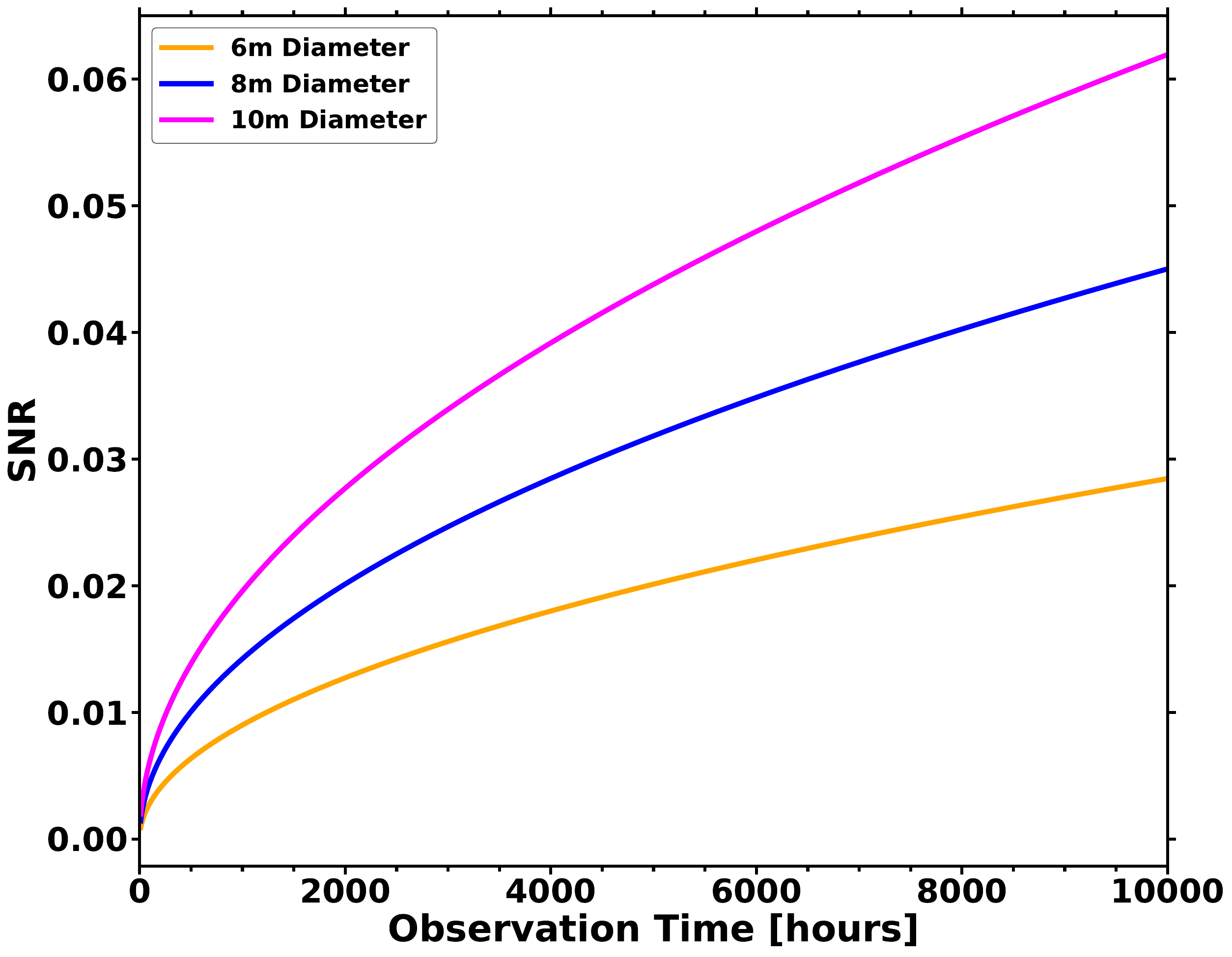}
       \begin{center}
             \textbf{(b)}
        \end{center}
        \end{minipage}
        
\caption{(a) Contrast difference with and without PAHs. The circles represent the wavelength bin points of the spectra. (b) Net signal-to-noise ratio (SNR) as a function of observation time for different telescope architectures (\textit{orange}: 6m, \textit{blue}: 8m, and \textit{magenta}: 10m). \textbf{A larger diameter is suitable to reach a higher SNR value with less observation time. However, with the current concentration of Naphthalene, Anthracene, Phenanthrene, and Pyrene on Earth, none of the studied telescope diameters can resolve collective UV spectral features with a significant SNR value.}}
\label{fig:telescope_architecture}
\end{figure*}

Figure \ref{fig:telescope_architecture}a depicts the PAH signal under varying telescope mirror diameters. An essential question arises: what constitutes the optimal scenario for detecting PAHs with a significant SNR value in this case? Figure \ref{fig:telescope_architecture}b presents SNR as a function of observation time, offering insights into this query. It becomes evident that a larger telescope collecting area is advantageous for PAH observation. \textbf{However, with the four molecules of our study, PAH detectability is not feasible given the limitations of the current instrument capabilities. However, it is important to note that these values represent a lower bound based on HWO, as the analysis is limited to the UV range (0.2–0.4 $\mathrm{\mu m}$) due to the constraints of PAH cross-sections and the consideration of only four PAHs from the broader PAH family.} Additionally, the native resolution of HWO in the UV band is too low to effectively resolve PAH features, which significantly increases the required observation time. A resolution of 30 to 50 in the UV range would be ideal for efficiently detecting PAH features. However, maintaining a low resolution in the UV domain is crucial for mitigating certain instrumental and observational limitations, which can enhance overall sensitivity and improve data quality (higher SNR).

\section{Discussion and Conclusion}
\label{sec:conclusion}

On Earth, PAHs are the result of industrial development and other human activities. Their detection in an exoplanet atmosphere could serve as evidence of extraterrestrial technology or life. Although PAH formation on hot Jupiters is still possible thermodynamically \citep{dubey2023polycyclic}, PAHs on Earth lack abiotic or non-anthropogenic sources. Our study focuses on the detectability of PAH signatures in the UV region (0.2-0.4 $\mathrm{\mu m}$) considering PAH concentrations at present-day Earth levels using future space mission Habitable Worlds Observatory (HWO). 0.3-0.4 $\mathrm{\mu m}$ wavelength band serves as a promising window to detect PAHs due to less overlap with other molecular cross-sections (see Figure \ref{fig:cross_section} (b)). We have limitations in going to the other wavelengths due to the unavailability of cross-sectional data for such heavy organic molecules. Some resources, such as the NASA Ames PAH database\footnote{\url{https://www.astrochemistry.org/pahdb/theoretical/3.20/default/browse}} \citep{bauschlicher2018nasa,mattioda2020nasa} offer extensive data on thousands of PAH transitions over a broad wavelength range, primarily focusing on the identification of PAHs within interstellar emission spectra. However, their implementation in exoplanetary atmospheres remains a challenge. 

One caveat of this study is that we assume PAH abundance to be constant throughout all pressure layers. Employing either an atmospheric chemistry model or a coupled chemistry-climate model would allow realistic prediction of PAH abundance in the planet's atmosphere. This will help us understand the sustainment of these heavy molecules inside the planet. Due to the unavailability of the PAH reaction network in the exoplanet community, more accurate calculation is hard to achieve at this stage. In addition, the analysis presented in this work does not include photochemical processes, which could impact the abundance of PAHs in planetary atmospheres. In Earth's atmosphere, PAHs are primarily destroyed through reactions with hydroxyl radicals (OH), ozone, and other oxidizing agents, resulting in relatively short atmospheric lifetimes, particularly under exposure to UV radiation \citep{arey2003photochemical, eagar2017impact}. These processes limit PAH survival, especially in the upper atmosphere, although PAHs may persist longer in the troposphere where UV exposure is lower and aerosol concentrations are higher. Larger PAHs have a different story. They are more resistant to these processes due to their structural resilience and lower reactivity, allowing them to persist longer in the troposphere. Therefore, while smaller PAHs may be rapidly depleted, larger, more complex PAHs are likely to remain in the atmosphere for extended periods. \textbf{Given the absence of complete cross-sectional data and a developed PAH reaction network for exoplanet atmospheres, we opted to present a scenario that assumes PAHs decay exponentially over time. With the lack of knowledge on PAH kinetics in an Earth-like atmosphere, this approach provides the best realistic scenario to calculate current Earth PAH concentrations, serving as a basis for further studies.} Incorporating detailed PAH loss mechanisms would offer a more complete understanding of PAH behavior, and we plan to address this in future work with a chemical kinetics model, once more data and reaction pathways become available.

\textbf{The limited scope of this study, focusing on four specific PAH species (Naphthalene, Anthracene, Phenanthrene, and Pyrene) highlights an inherent constraint in current observational analyses of PAHs. While these four molecules were assumed to have atmospheric lifetimes of a few days, their actual persistence is more complex. Complete degradation in the troposphere region requires intense UV radiation, which is not prevalent under typical atmospheric conditions. Instead, their lifetimes are governed by reactions with reactive radicals as mentioned above, leading to the formation of substituted PAHs rather than their complete removal from the atmosphere. As a result, PAHs remain in the atmospheric system in various stable forms, contributing to their long-term presence. This persistence underscores the need for future studies to broaden the scope by including a more comprehensive range of PAH species. The PAH family exhibits diverse chemical structures and behaviors, and studying them collectively is crucial for accurately identifying their spectral signatures and understanding their detectability. Restricting the analysis to only a few species provides an incomplete picture, potentially underestimating their overall contribution and presence in the atmosphere. Expanding the analysis to cover a wider spectrum of PAHs would enable a more robust assessment of their detectability and provide deeper insights into their roles in atmospheric chemistry and observational signatures.}

\textbf{The UV spectral signatures of PAHs are highly robust due to the stability of their delocalized $\mathrm{\pi}$-electron systems, as demonstrated by the strong $\mathrm{\pi}$ $\rightarrow$ $\mathrm{\pi^*}$ electronic transitions that dominate their absorption spectra (with primary absorption bands appearing around 200–250 nm and weaker bands in the 250–300 nm range) \citep{malloci2004electronic,halasinski2005investigation}. While the presence of electron-donating or electron-withdrawing substituents can induce shifts in the primary absorption bands, these shifts do not significantly alter the distinctiveness or detectability of the UV features of PAHs. This robustness suggests that even substituted PAHs retain spectral characteristics that make them suitable for detection in atmospheric and astrophysical environments. While the current study with the decay term provides a conservative estimate, it also highlights a limitation: by focusing solely on the parent PAH molecules, we may overlook the broader contribution of the PAH family, including substituted species. Given that the substitution minimally impacts the UV spectral features, future studies should consider the entire PAH family without applying decay terms to improve the robustness of their detection in UV observations.}

To mitigate the issue of varying PAH abundances across atmospheric layers, we opted for reflectance spectra, which are sensitive to the total column density of PAHs. Therefore, a reduction in PAH abundance above the troposphere may not significantly impact the reflectance spectra, unlike in transmission spectra, where upper atmospheric abundances play a more crucial role. Another aspect to consider is whether investigating the detectability of PAHs on planets with higher concentrations is a reasonable endeavor. Due to their role in human health hazards including cancer, their production has substantially been reduced by different government policies worldwide. Additionally, the industrial revolution in the past few decades has contributed to a decrease in PAH emissions into the Earth's atmosphere. Similar developments may happen to extraterrestrial civilizations as well. However, the effectiveness of these reductions is highly contingent upon factors such as the specific climate requirements of the affected species, the atmospheric composition of the planet, and numerous assumptions regarding the long-term goals and coordination efforts of the involved entities.

In this work, we predict the detection of PAHs using the proposed HWO telescope architecture. \textbf{Even with the molecular concentrations of Naphthalene, Anthracene, Phenanthrene, and Pyrene scaled by factors of 5 and 10 relative to the current Earth's levels, achieving a significant SNR remains impossible.} The results could be substantially improved by addressing two main aspects: (1) acquiring cross-sections for the entire wavelength range for Naphthalene, Anthracene, Phenanthrene, and Pyrene and (2) accounting for other PAHs that are prevalent in the atmosphere but currently lack measured cross-sections. The combined impact of all PAHs on planetary spectra could thus establish PAHs as a potential technosignature from the perspective of Earth. 

However, a better sensitivity can be achieved through alternative telescope architectures with larger mirror sizes. Ground-based observatories, with their larger collecting areas, complement the future HWO mission by covering visible and NIR bands. If PAH cross-sections were available across HWO's full wavelength range (0.2-2 $\mathrm{\mu m}$), ground-based observations could offer valuable complementary analysis, especially in the overlapping coverage starting in the optical bands (0.4 $\mathrm{\mu m}$ or longer). From the perspective of space telescopes, a 15-meter telescope offers improved sensitivity with its larger architecture. Taking into account the cumulative PAH concentration levels over the last six decades and the uncertainty in the measurement (twice the measured concentration), we investigate various telescope mirror diameters to understand sensitivity towards PAH observation better. Optimal scenarios involve larger diameters (a 10m mirror diameter). \textbf{At current Earth concentrations of the four molecules, even a 10m diameter telescope fails to resolve their collective spectral features with a decent signal-to-noise ratio, even after 10000 hours of observation time.} It is important to note that the UV band resolution of HWO is insufficient for properly resolving PAH signatures. The lower resolution fails to capture all the necessary wavelength bin points, resulting in a significantly reduced signal strength for PAHs — approximately 2.5 times lower than what would be obtained with a resolution of 40. While a slightly higher resolution could be advantageous, we are constrained here to a lower resolution due to various instrumental and observational limitations in the UV band.

The presence of PAHs on Earth predominantly arises from biomass burning. Since biomass is linked to oxygenic photosynthesis, the detection of PAHs (like other industrially processed compounds such as CFCs, $\mathrm{NO_2}$ etc.) suggests the existence of an oxygen-rich atmosphere on the planet, indicative of vegetation presence as well \cite{balbi2024oxygen}. Oxygen as a potential biosignature has also been widely explored in the literature \citep{schwieterman2018exoplanet}. The prevalence of specific PAHs produced on Earth remains uncertain in extraterrestrial environments, even among civilizations with similar industrial processes. The diversity within the PAH family is extensive (see NASA Ames PAH database) necessitating further investigation to assess the detectability of a broader class of PAHs.  However, during the study of exo-Earth candidates for potential biosignatures, an additional requirement of integrated observing time may be required at this stage to detect current Earth-level PAHs. Obtaining measured cross-sections for the PAH class has become a necessity for gaining a deeper understanding of their presence in the atmospheres. Clouds were not considered in this study, although PAH cross-sections overlap with similar wavelength regions as $\mathrm{NO_2}$. The presence of clouds and aerosols can diminish detectability and potentially mimic $\mathrm{NO_2}$ features \citep{kopparapu2021nitrogen}, making it difficult to uniquely distinguish between PAH and $\mathrm{NO_2}$ signatures. Exploring the atmospheres of planets, particularly those with clouds, around other stars offers the potential to refine our understanding of PAH detectability and presents promising prospects for future study.


\begin{acknowledgments}
\textit{Acknowledgments}: R. K gratefully acknowledges support from the NASA Exobiology program under grant
21-EXO21-0025. We would like to thank the anonymous referees for constructive comments that helped improve the manuscript. 
\end{acknowledgments}

\bibliography{references}{}

\begin{thebibliography}{}
\expandafter\ifx\csname natexlab\endcsname\relax\def\natexlab#1{#1}\fi
\providecommand{\url}[1]{\href{#1}{#1}}
\providecommand{\dodoi}[1]{doi:~\href{http://doi.org/#1}{\nolinkurl{#1}}}
\providecommand{\doeprint}[1]{\href{http://ascl.net/#1}{\nolinkurl{http://ascl.net/#1}}}
\providecommand{\doarXiv}[1]{\href{https://arxiv.org/abs/#1}{\nolinkurl{https://arxiv.org/abs/#1}}}

\bibitem[{ost(1987)}]{osti_5473300}
 1987.
\newblock \url{https://www.osti.gov/biblio/5473300}

\bibitem[{Arey \& Atkinson(2003)}]{arey2003photochemical}
Arey, J., \& Atkinson, R. 2003, PAHs: An ecotoxicological perspective, 47

\bibitem[{Arnold(2005)}]{arnold2005transit}
Arnold, L.~F. 2005, The Astrophysical Journal, 627, 534

\bibitem[{Balakrishnan {et~al.}(2011)Balakrishnan, Ramaswamy, Sambandam, Thangavel, Ghosh, Johnson, Mukhopadhyay, Venugopal, \& Thanasekaraan}]{balakrishnan2011air}
Balakrishnan, K., Ramaswamy, P., Sambandam, S., {et~al.} 2011, Global health action, 4, 5638

\bibitem[{Balbi \& Frank(2024)}]{balbi2024oxygen}
Balbi, A., \& Frank, A. 2024, Nature Astronomy, 8, 39

\bibitem[{Bauschlicher {et~al.}(2018)Bauschlicher, Ricca, Boersma, \& Allamandola}]{bauschlicher2018nasa}
Bauschlicher, C.~W., Ricca, A., Boersma, C., \& Allamandola, L. 2018, The Astrophysical Journal Supplement Series, 234, 32

\bibitem[{{Benneke} {et~al.}(2019){Benneke}, {Wong}, {Piaulet}, {Knutson}, {Lothringer}, {Morley}, {Crossfield}, {Gao}, {Greene}, {Dressing}, {Dragomir}, {Howard}, {McCullough}, {Kempton}, {Fortney}, \& {Fraine}}]{2019ApJ...887L..14B}
{Benneke}, B., {Wong}, I., {Piaulet}, C., {et~al.} 2019, \apjl, 887, L14, \dodoi{10.3847/2041-8213/ab59dc}

\bibitem[{Bracewell(1960)}]{bracewell1960communications}
Bracewell, R.~N. 1960, Nature, 186, 670

\bibitem[{Cadieux {et~al.}(2024)Cadieux, Doyon, MacDonald, Turbet, Artigau, Lim, Radica, Fauchez, Salhi, Dang, {et~al.}}]{cadieux2024transmission}
Cadieux, C., Doyon, R., MacDonald, R.~J., {et~al.} 2024, The Astrophysical Journal Letters, 970, L2

\bibitem[{Campbell(2005)}]{Campbell_2005}
Campbell, J.~B. 2005, Proceedings of the International Astronomical Union, 1, 247–250, \dodoi{10.1017/S1743921306009392}

\bibitem[{{Carrigan}(2009)}]{2009ASPC..420..415C}
{Carrigan}, R.~A., J. 2009, in Astronomical Society of the Pacific Conference Series, Vol. 420, Bioastronomy 2007: Molecules, Microbes and Extraterrestrial Life, ed. K.~J. {Meech}, J.~V. {Keane}, M.~J. {Mumma}, J.~L. {Siefert}, \& D.~J. {Werthimer}, 415

\bibitem[{{Catling} {et~al.}(2018){Catling}, {Krissansen-Totton}, {Kiang}, {Crisp}, {Robinson}, {DasSarma}, {Rushby}, {Del Genio}, {Bains}, \& {Domagal-Goldman}}]{2018AsBio..18..709C}
{Catling}, D.~C., {Krissansen-Totton}, J., {Kiang}, N.~Y., {et~al.} 2018, Astrobiology, 18, 709, \dodoi{10.1089/ast.2017.1737}

\bibitem[{Charbonneau {et~al.}(2002)Charbonneau, Brown, Noyes, \& Gilliland}]{Charbonneau_2002}
Charbonneau, D., Brown, T.~M., Noyes, R.~W., \& Gilliland, R.~L. 2002, The Astrophysical Journal, 568, 377, \dodoi{10.1086/338770}

\bibitem[{Closs {et~al.}(2020)Closs, Fuks, Bechtel, \& Trapp}]{closs2020prebiotically}
Closs, A.~C., Fuks, E., Bechtel, M., \& Trapp, O. 2020, Chemistry--A European Journal, 26, 10702

\bibitem[{{Crossfield}(2015)}]{2015PASP..127..941C}
{Crossfield}, I. J.~M. 2015, \pasp, 127, 941, \dodoi{10.1086/683115}

\bibitem[{Dinelli {et~al.}(2013)Dinelli, L{\'o}pez-Puertas, Adriani, Moriconi, Funke, Garc{\'\i}a-Comas, \& D'Aversa}]{dinelli2013unidentified}
Dinelli, B., L{\'o}pez-Puertas, M., Adriani, A., {et~al.} 2013, Geophysical Research Letters, 40, 1489

\bibitem[{Draine \& Li(2007)}]{draine2007infrared}
Draine, B., \& Li, A. 2007, The Astrophysical Journal, 657, 810

\bibitem[{Dubey {et~al.}(2023)Dubey, Gr{\"u}bel, Arenales-Lope, Molaverdikhani, Ercolano, Rab, \& Trapp}]{dubey2023polycyclic}
Dubey, D., Gr{\"u}bel, F., Arenales-Lope, R., {et~al.} 2023, Astronomy \& Astrophysics, 678, A53

\bibitem[{Dyson(1960)}]{dyson1960search}
Dyson, F.~J. 1960, Science, 131, 1667

\bibitem[{Eagar {et~al.}(2017)Eagar, Ervens, \& Herckes}]{eagar2017impact}
Eagar, J.~D., Ervens, B., \& Herckes, P. 2017, Atmospheric Environment, 160, 132

\bibitem[{Ehrenfreund \& Charnley(2000)}]{ehrenfreund2000organic}
Ehrenfreund, P., \& Charnley, S.~B. 2000, Annual Review of Astronomy and Astrophysics, 38, 427

\bibitem[{Ehrenfreund {et~al.}(2006)Ehrenfreund, Rasmussen, Cleaves, \& Chen}]{ehrenfreund2006experimentally}
Ehrenfreund, P., Rasmussen, S., Cleaves, J., \& Chen, L. 2006, Astrobiology, 6, 490

\bibitem[{Ehrenfreund {et~al.}(2007)Ehrenfreund, Ruiterkamp, Peeters, Foing, Salama, \& Martins}]{ehrenfreund2007organics}
Ehrenfreund, P., Ruiterkamp, R., Peeters, Z., {et~al.} 2007, Planetary and Space Science, 55, 383

\bibitem[{El-Masri(2005)}]{el2005toxicological}
El-Masri, H. 2005, Toxicological profile for naphthalene, 1-methylnaphthalene, and 2-methylnaphthalene (US Department of Health and Human Services, Public Health Service, Agency~…)

\bibitem[{Ercolano {et~al.}(2022)Ercolano, Rab, Molaverdikhani, Edwards, Preibisch, Testi, Kamp, \& Thi}]{ercolano2022observations}
Ercolano, B., Rab, C., Molaverdikhani, K., {et~al.} 2022, Monthly Notices of the Royal Astronomical Society, 512, 430

\bibitem[{{Forgan}(2013)}]{2013JBIS...66..144F}
{Forgan}, D.~H. 2013, Journal of the British Interplanetary Society, 66, 144, \dodoi{10.48550/arXiv.1306.1672}

\bibitem[{Freitas~Jr \& Valdes(1980)}]{freitas1980search}
Freitas~Jr, R.~A., \& Valdes, F. 1980, Icarus, 42, 442

\bibitem[{Fujii {et~al.}(2018)Fujii, Angerhausen, Deitrick, Domagal-Goldman, Grenfell, Hori, Kane, Pall{\'e}, Rauer, Siegler, {et~al.}}]{fujii2018exoplanet}
Fujii, Y., Angerhausen, D., Deitrick, R., {et~al.} 2018, Astrobiology, 18, 739

\bibitem[{Gorti {et~al.}(2009)Gorti, Dullemond, \& Hollenbach}]{gorti2009time}
Gorti, U., Dullemond, C., \& Hollenbach, D. 2009, The Astrophysical Journal, 705, 1237

\bibitem[{Greene {et~al.}(2023)Greene, Bell, Ducrot, Dyrek, Lagage, \& Fortney}]{greene2023thermal}
Greene, T.~P., Bell, T.~J., Ducrot, E., {et~al.} 2023, Nature, 618, 39

\bibitem[{Grenfell(2017)}]{grenfell2017review}
Grenfell, J.~L. 2017, Physics Reports, 713, 1

\bibitem[{Grosch {et~al.}(2015)Grosch, S{\'a}rossy, Egsgaard, \& Fateev}]{grosch2015uv}
Grosch, H., S{\'a}rossy, Z., Egsgaard, H., \& Fateev, A. 2015, Journal of Quantitative Spectroscopy and Radiative Transfer, 156, 17

\bibitem[{Halasinski {et~al.}(2005)Halasinski, Salama, \& Allamandola}]{halasinski2005investigation}
Halasinski, T., Salama, F., \& Allamandola, L. 2005, The Astrophysical Journal, 628, 555

\bibitem[{Haqq-Misra {et~al.}(2022)Haqq-Misra, Kopparapu, Fauchez, Frank, Wright, \& Lingam}]{haqq2022detectability}
Haqq-Misra, J., Kopparapu, R., Fauchez, T.~J., {et~al.} 2022, The Planetary Science Journal, 3, 60

\bibitem[{Haqq-Misra \& Kopparapu(2012)}]{haqq2012likelihood}
Haqq-Misra, J., \& Kopparapu, R.~K. 2012, Acta Astronautica, 72, 15

\bibitem[{Haqq-Misra {et~al.}(2020)Haqq-Misra, Kopparapu, \& Schwieterman}]{haqq2020observational}
Haqq-Misra, J., Kopparapu, R.~K., \& Schwieterman, E. 2020, Astrobiology, 20, 572

\bibitem[{Joblin \& Tielens(2011)}]{joblin2011pahs}
Joblin, C., \& Tielens, A. G. G.~M. 2011, EAS publications series, 46

\bibitem[{Kaltenegger(2017)}]{kaltenegger2017characterize}
Kaltenegger, L. 2017, Annual Review of Astronomy and Astrophysics, 55, 433

\bibitem[{{Kim} {et~al.}(2012){Kim}, {Im}, {Lee}, {Lee}, {Jun}, {Nakagawa}, {Matsuhara}, {Wada}, {Oyabu}, {Takagi}, {Inami}, {Ohyama}, {Yamada}, {Helou}, {Armus}, \& {Shi}}]{2012ApJ...760..120K}
{Kim}, J.~H., {Im}, M., {Lee}, H.~M., {et~al.} 2012, \apj, 760, 120, \dodoi{10.1088/0004-637X/760/2/120}

\bibitem[{Kipping \& Teachey(2016)}]{kipping2016cloaking}
Kipping, D.~M., \& Teachey, A. 2016, Monthly Notices of the Royal Astronomical Society, 459, 1233

\bibitem[{Kitagawa(1968)}]{kitagawa1968absorption}
Kitagawa, T. 1968, Journal of Molecular Spectroscopy, 26, 1

\bibitem[{Kopparapu {et~al.}(2021)Kopparapu, Arney, Haqq-Misra, Lustig-Yaeger, \& Villanueva}]{kopparapu2021nitrogen}
Kopparapu, R., Arney, G., Haqq-Misra, J., Lustig-Yaeger, J., \& Villanueva, G. 2021, The Astrophysical Journal, 908, 164

\bibitem[{Kuhn \& Berdyugina(2015)}]{kuhn2015global}
Kuhn, J.~R., \& Berdyugina, S.~V. 2015, International journal of astrobiology, 14, 401

\bibitem[{Laflamme \& Hites(1978)}]{laflamme1978global}
Laflamme, R.~E., \& Hites, R.~A. 1978, Geochimica et cosmochimica Acta, 42, 289

\bibitem[{Lammer {et~al.}(2019)Lammer, Spro{\ss}, Grenfell, Scherf, Fossati, Lendl, \& Cubillos}]{lammer2019role}
Lammer, H., Spro{\ss}, L., Grenfell, J.~L., {et~al.} 2019, Astrobiology, 19, 927

\bibitem[{Lim {et~al.}(2023)Lim, Benneke, Doyon, MacDonald, Piaulet, Artigau, Coulombe, Radica, L’Heureux, Albert, {et~al.}}]{lim2023atmospheric}
Lim, O., Benneke, B., Doyon, R., {et~al.} 2023, The Astrophysical Journal Letters, 955, L22

\bibitem[{{Lin} {et~al.}(2014){Lin}, {Gonzalez Abad}, \& {Loeb}}]{2014ApJ...792L...7L}
{Lin}, H.~W., {Gonzalez Abad}, G., \& {Loeb}, A. 2014, \apjl, 792, L7, \dodoi{10.1088/2041-8205/792/1/L7}

\bibitem[{Lingam \& Loeb(2017)}]{lingam2017natural}
Lingam, M., \& Loeb, A. 2017, Monthly Notices of the Royal Astronomical Society: Letters, 470, L82

\bibitem[{Lingam \& Loeb(2021)}]{lingam2021life}
---. 2021, Life in the cosmos: From biosignatures to technosignatures (Harvard University Press)

\bibitem[{Loeb \& Turner(2012)}]{loeb2012detection}
Loeb, A., \& Turner, E.~L. 2012, Astrobiology, 12, 290

\bibitem[{L{\'o}pez-Puertas {et~al.}(2013)L{\'o}pez-Puertas, Dinelli, Adriani, Funke, Garc{\'\i}a-Comas, Moriconi, D’Aversa, Boersma, \& Allamandola}]{lopez2013large}
L{\'o}pez-Puertas, M., Dinelli, B.~M., Adriani, A., {et~al.} 2013, The Astrophysical Journal, 770, 132

\bibitem[{Lustig-Yaeger {et~al.}(2019)Lustig-Yaeger, Meadows, \& Lincowski}]{Lustig-Yaeger_2019}
Lustig-Yaeger, J., Meadows, V.~S., \& Lincowski, A.~P. 2019, The Astronomical Journal, 158, 27, \dodoi{10.3847/1538-3881/ab21e0}

\bibitem[{Lustig-Yaeger {et~al.}(2023)Lustig-Yaeger, Fu, May, Ceballos, Moran, Peacock, Stevenson, Kirk, L{\'o}pez-Morales, MacDonald, {et~al.}}]{lustig2023jwst}
Lustig-Yaeger, J., Fu, G., May, E., {et~al.} 2023, Nature Astronomy, 7, 1317

\bibitem[{Malloci {et~al.}(2004)Malloci, Mulas, \& Joblin}]{malloci2004electronic}
Malloci, G., Mulas, G., \& Joblin, C. 2004, Astronomy \& Astrophysics, 426, 105

\bibitem[{Mattioda {et~al.}(2020)Mattioda, Hudgins, Boersma, Bauschlicher, Ricca, Cami, Peeters, de~Armas, Saborido, \& Allamandola}]{mattioda2020nasa}
Mattioda, A., Hudgins, D., Boersma, C., {et~al.} 2020, The Astrophysical Journal Supplement Series, 251, 22

\bibitem[{Meadows {et~al.}(2018)Meadows, Reinhard, Arney, Parenteau, Schwieterman, Domagal-Goldman, Lincowski, Stapelfeldt, Rauer, DasSarma, {et~al.}}]{meadows2018exoplanet}
Meadows, V.~S., Reinhard, C.~T., Arney, G.~N., {et~al.} 2018, Astrobiology, 18, 630

\bibitem[{NRC(1983)}]{national1983atmospheric}
NRC. 1983, in Polycyclic Aromatic Hydrocarbons: Evaluation of Sources and Effects (National Academies Press (US))

\bibitem[{Owen(1980)}]{owen1980search}
Owen, T. 1980, in Strategies for the Search for Life in the Universe: A Joint Session of Commissions 16, 40, and 44, Held in Montreal, Canada, During the IAU General Assembly, 15 and 16 August, 1979 (Springer), 177--185

\bibitem[{O’Malley-James \& Kaltenegger(2019)}]{o2019expanding}
O’Malley-James, J.~T., \& Kaltenegger, L. 2019, The Astrophysical Journal Letters, 879, L20

\bibitem[{Participants(2018)}]{participants2018nasa}
Participants, N. 2018, arXiv preprint arXiv:1812.08681

\bibitem[{Perera(1997)}]{perera1997environment}
Perera, F.~P. 1997, Science, 278, 1068

\bibitem[{Puzzarini {et~al.}(2017)Puzzarini, Baiardi, Bloino, Barone, Murphy, Drew, \& Ali}]{puzzarini2017spectroscopic}
Puzzarini, C., Baiardi, A., Bloino, J., {et~al.} 2017, The Astronomical Journal, 154, 82

\bibitem[{Rapacioli {et~al.}(2006)Rapacioli, Calvo, Joblin, Parneix, Toublanc, \& Spiegelman}]{rapacioli2006formation}
Rapacioli, M., Calvo, F., Joblin, C., {et~al.} 2006, Astronomy \& Astrophysics, 460, 519

\bibitem[{Ravindra {et~al.}(2008)Ravindra, Sokhi, \& Van~Grieken}]{ravindra2008atmospheric}
Ravindra, K., Sokhi, R., \& Van~Grieken, R. 2008, Atmospheric environment, 42, 2895

\bibitem[{Rose \& Wright(2004)}]{rose2004inscribed}
Rose, C., \& Wright, G. 2004, Nature, 431, 47

\bibitem[{Sandstrom {et~al.}(2011)Sandstrom, Bolatto, Bot, Draine, Ingalls, Israel, Jackson, Leroy, Li, Rubio, {et~al.}}]{sandstrom2011spitzer}
Sandstrom, K.~M., Bolatto, A.~D., Bot, C., {et~al.} 2011, The Astrophysical Journal, 744, 20

\bibitem[{Schneider {et~al.}(2010)Schneider, L{\'e}ger, Fridlund, White, Eiroa, Henning, Herbst, Lammer, Liseau, Paresce, {et~al.}}]{schneider2010far}
Schneider, J., L{\'e}ger, A., Fridlund, M., {et~al.} 2010, Astrobiology, 10, 121

\bibitem[{Schwieterman {et~al.}(2018)Schwieterman, Kiang, Parenteau, Harman, DasSarma, Fisher, Arney, Hartnett, Reinhard, Olson, {et~al.}}]{schwieterman2018exoplanet}
Schwieterman, E.~W., Kiang, N.~Y., Parenteau, M.~N., {et~al.} 2018, Astrobiology, 18, 663

\bibitem[{Seager {et~al.}(2012)Seager, Schrenk, \& Bains}]{seager2012astrophysical}
Seager, S., Schrenk, M., \& Bains, W. 2012, Astrobiology, 12, 61

\bibitem[{Shahsavar {et~al.}(2023)Shahsavar, Zahedi, Shiroudi, \& Chahkandi}]{shahsavar2023atmospheric}
Shahsavar, F., Zahedi, E., Shiroudi, A., \& Chahkandi, B. 2023, Journal of Molecular Graphics and Modelling, 121, 108426

\bibitem[{Shen {et~al.}(2013)Shen, Huang, Wang, Zhu, Li, Shen, Wang, Zhang, Chen, Lu, {et~al.}}]{shen2013global}
Shen, H., Huang, Y., Wang, R., {et~al.} 2013, Environmental science \& technology, 47, 6415

\bibitem[{Shklovsky \& Sagan(1966)}]{shklovsky1966intelligent}
Shklovsky, I.~S., \& Sagan, C. 1966, New York: Delta, 10

\bibitem[{Stevens {et~al.}(2016)Stevens, Forgan, \& James}]{stevens2016observational}
Stevens, A., Forgan, D., \& James, J.~O. 2016, International Journal of Astrobiology, 15, 333

\bibitem[{Suto {et~al.}(1992)Suto, Wang, Shan, \& Lee}]{suto1992quantitative}
Suto, M., Wang, X., Shan, J., \& Lee, L. 1992, Journal of Quantitative Spectroscopy and Radiative Transfer, 48, 79

\bibitem[{Tabor \& Loeb(2021)}]{tabor2021detectability}
Tabor, E., \& Loeb, A. 2021, arXiv preprint arXiv:2105.08081

\bibitem[{Tarter(2006)}]{tarter2006evolution}
Tarter, J.~C. 2006, Proceedings of the International Astronomical Union, 2, 14

\bibitem[{Thi {et~al.}(2019)Thi, Lesur, Woitke, Kamp, Rab, \& Carmona}]{thi2019radiation}
Thi, W., Lesur, G., Woitke, P., {et~al.} 2019, Astronomy \& Astrophysics, 632, A44

\bibitem[{Th{\"o}ny \& Rossi(1997)}]{thony1997gas}
Th{\"o}ny, A., \& Rossi, M.~J. 1997, Journal of Photochemistry and Photobiology A: Chemistry, 104, 25

\bibitem[{Tielens(2008)}]{tielens2008interstellar}
Tielens, A.~G. 2008, Annu. Rev. Astron. Astrophys., 46, 289

\bibitem[{{Tsiaras} {et~al.}(2019){Tsiaras}, {Waldmann}, {Tinetti}, {Tennyson}, \& {Yurchenko}}]{2019NatAs...3.1086T}
{Tsiaras}, A., {Waldmann}, I.~P., {Tinetti}, G., {Tennyson}, J., \& {Yurchenko}, S.~N. 2019, Nature Astronomy, 3, 1086, \dodoi{10.1038/s41550-019-0878-9}

\bibitem[{Villanueva {et~al.}(2022)Villanueva, Liuzzi, Faggi, Protopapa, Kofman, Fauchez, Stone, \& Mandell}]{villanueva2022fundamentals}
Villanueva, G.~L., Liuzzi, G., Faggi, S., {et~al.} 2022, Fundamentals of the Planetary Spectrum Generator. 2022 edition of the handbook by GL Villanueva et al. ISBN 978-0-578-36143-7

\bibitem[{Villanueva {et~al.}(2018)Villanueva, Smith, Protopapa, Faggi, \& Mandell}]{villanueva2018planetary}
Villanueva, G.~L., Smith, M.~D., Protopapa, S., Faggi, S., \& Mandell, A.~M. 2018, Journal of Quantitative Spectroscopy and Radiative Transfer, 217, 86

\bibitem[{Wakelam \& Herbst(2008)}]{wakelam2008polycyclic}
Wakelam, V., \& Herbst, E. 2008, The Astrophysical Journal, 680, 371

\bibitem[{Walker {et~al.}(2018)Walker, Bains, Cronin, DasSarma, Danielache, Domagal-Goldman, Kacar, Kiang, Lenardic, Reinhard, {et~al.}}]{walker2018exoplanet}
Walker, S.~I., Bains, W., Cronin, L., {et~al.} 2018, Astrobiology, 18, 779

\bibitem[{Wang {et~al.}(2013)Wang, Tao, Ciais, Shen, Huang, Chen, Shen, Wang, Li, Zhang, {et~al.}}]{wang2013high}
Wang, R., Tao, S., Ciais, P., {et~al.} 2013, Atmospheric Chemistry and Physics, 13, 5189

\bibitem[{Whitmire \& Wright(1980)}]{whitmire1980nuclear}
Whitmire, D.~P., \& Wright, D.~P. 1980, Icarus, 42, 149

\bibitem[{Wright {et~al.}(2014)Wright, Griffith, Sigurdsson, Povich, \& Mullan}]{wright2014g}
Wright, J., Griffith, R., Sigurdsson, S., Povich, M., \& Mullan, B. 2014, The Astrophysical Journal, 792, 27

\bibitem[{Wright {et~al.}(2019)Wright, Allen, Alvarado-G{\' o}mez, Angerhausen, Apai, Atri, Balbi, Barclay, Barentsen, Beasley, Beatty, Behmard, Berea, Boyajian, Bridge, Bryson, Bytof, Cleaves, Colon, Cordes, Cowing, Curtis, Davenport, Davies, DeMarines, Denning, Dick, Dong, Dutil, Edmonds, Enriquez, Eubanks, Fernandez, Frank, Torre, Gajjar, Garrett, Gelino, Geller, Giles, Gillum, G{\' o}mez, Graham, Grimaldi, Grinspoon, Haqq-Misra, Hellbourg, Helman, Horowitz, Howard, Isaacson, Jackson, Jia, Kainu, Kanodia, Kawaler, Kendall, Khan, Kipping, Kite, Knuth, Kopparapu, Korpela, Laine, Lau, Lesyna, Loureiro, MacDonald, Margot, Mendez, Mishra, Morrison, Mullally, Mullan, Nita, O'Neil, Pass, Paulino-Lima, Piotelat, Pope, Raymond, Ricker, Riley, Robertson, Rocha, Rodriguez, Rosenthal, Roy, Rybarczyk, Souza, Sallmen, Santander, Scharf, Schwieterman, Seiler, Shabram, Sheikh, Shostak, Shrestha, Siemion, Sigurdsson, Sneed, Socas-Navarro, Soderblom, Solmaz, Subramaniam, Suresh, Tan, Tanner, Tarter, Taylor, Terrien, Turner,
  Vakoch, Villa, Walkowicz, Wang, Weiss, Werthimer, Williams, Winn, Wishnow, Worden, Wright, Zackrisson, Zanis, \& Zarka}]{Wright2019Searches}
Wright, J., Allen, V., Alvarado-G{\' o}mez, J.~D., {et~al.} 2019, Bulletin of the AAS, 51

\bibitem[{Wright {et~al.}(2015)Wright, Cartier, Zhao, Jontof-Hutter, \& Ford}]{wright2015g}
Wright, J.~T., Cartier, K.~M., Zhao, M., Jontof-Hutter, D., \& Ford, E.~B. 2015, The Astrophysical Journal, 816, 17

\bibitem[{Wright {et~al.}(2018)Wright, Sheikh, Alm{\'a}r, Denning, Dick, \& Tarter}]{wright2018recommendations}
Wright, J.~T., Sheikh, S., Alm{\'a}r, I., {et~al.} 2018, arXiv preprint arXiv:1809.06857

\bibitem[{Zhao {et~al.}(2016)Zhao, Zhang, \& Wang}]{zhao2016atmospheric}
Zhao, N., Zhang, Q., \& Wang, W. 2016, Science of the Total Environment, 563, 1008

\bibitem[{Zhou {et~al.}(2017)Zhou, Xing, Lang, Chen, Cheng, Wei, Wei, \& Liu}]{zhou2017comprehensive}
Zhou, Y., Xing, X., Lang, J., {et~al.} 2017, Atmospheric Chemistry and Physics, 17, 2839

\end{thebibliography}
\bibliographystyle{aasjournal}


\appendix
\section{Cross-sectional requirement for PAHs}
\label{sec:future}

As previously highlighted, a significant limitation arises from the absence of comprehensive PAH cross-sectional data. Given the considerable variability in both species composition and their concentrations worldwide, there is a pressing need for additional cross-sectional data to enhance the fidelity of atmospheric modeling. Such data are essential for assessing the detectability of aromatic hydrocarbon entities within the Earth's atmosphere. This molecular category displays similar cross-sectional features in the IR region, with notable peaks at approximately 3.3 $\mathrm{\mu m}$ and a range of peaks between 6 and 12 $\mathrm{\mu m}$, albeit with varying intensities. This underscores the necessity of acquiring cross-sectional data over a broader wavelength spectrum, encompassing both UV and IR regions, to fully characterize these molecular signatures. Such comprehensive data would not only amplify the combined signal strengths originating from PAHs but also streamline observational processes, thereby advancing our ability to detect these compounds effectively. We acknowledge that the spectral features discussed, such as the 3.3 $\mathrm{\mu m}$ C-H stretching band, are also characteristic of simpler hydrocarbons like $\mathrm{CH_4}$, $\mathrm{C_2H_2}$, $\mathrm{C_2H_4}$, $\mathrm{C_2H_6}$, and so on. This spectral overlap presents a challenge in distinguishing PAHs from these other hydrocarbons in exoplanetary atmospheres. Future work should aim to develop more precise cross-sectional data and advanced observational techniques to differentiate PAHs from these overlapping spectral features. In this context, we propose specific molecules whose cross-sectional data (in the wavelength range 0.2 - 12 $\mathrm{\mu m}$) are pivotal due to their significantly higher abundance in the Earth's atmosphere: Naphthalene ($\mathrm{C_{10}H_{8}}$), Anthracene ($\mathrm{C_{14}H_{10}}$), Phenanthrene ($\mathrm{C_{14}H_{10}}$), Pyrene ($\mathrm{C_{16}H_{10}}$), Fluorene ($\mathrm{C_{13}H_{10}}$), Fluorantheene ($\mathrm{C_{16}H_{10}}$), Acenaphthylene ($\mathrm{C_{16}H_{10}}$), Acenaphthene ($\mathrm{C_{12}H_{10}}$), Perylene ($\mathrm{C_{20}H_{12}}$), Coronene ($\mathrm{C_{24}H_{12}}$), Ovalene ($\mathrm{C_{32}H_{14}}$), Benzo[b]fluoranthene ($\mathrm{C_{20}H_{12}}$), Benzo[k]fluoranthene ($\mathrm{C_{20}H_{12}}$), 2-methylnaphthalene ($\mathrm{C_{11}H_{10}}$), and 1-methylnaphthalene ($\mathrm{C_{11}H_{10}}$). Furthermore, their availability can significantly aid in the investigation of PAH signatures during exoplanet observations.




\end{document}